\documentclass[acmsmall]{acmart}

\usepackage{natbib}
\usepackage{multirow}
\usepackage{microtype}
\usepackage{subfigure}
\usepackage{makecell}
\usepackage{color}
\usepackage{colortbl}
\usepackage{bm}
\usepackage{bbm}
\usepackage{cancel}
\usepackage[
  font = small,
  tableposition = top
]{caption}

\usepackage{algorithm}
\usepackage{algpseudocode}
\usepackage{booktabs}

\usepackage{multicol}
\usepackage{amsmath}
\usepackage{enumitem}

\usepackage{tikz}

\usepackage{tcolorbox}
\usepackage{threeparttable}
\usepackage[normalem]{ulem}

\usepackage{xcolor}
\usepackage{lineno}

\newcommand{\D}[1]{}

\newcommand{\ours}{\textsc{PuzzleMark}}

\newcommand{\graycell}{\cellcolor[rgb]{0.871,0.871,0.871}}

\AtBeginDocument{%
  }

\setcopyright{cc}
\setcctype{by}
\acmDOI{10.1145/3797122}
\acmYear{2026}
\acmJournal{PACMSE}
\acmVolume{3}
\acmNumber{FSE}
\acmArticle{FSE027}
\acmMonth{7}
\acmSubmissionID{fse26maina-p509-p}
\received{2025-09-12}
\received[accepted]{2025-12-22}

\begin{document}

\title[PuzzleMark: Implicit Jigsaw Learning for Robust Code Dataset Watermarking in Neural Code Completion \dots]
{PuzzleMark: Implicit Jigsaw Learning for Robust Code Dataset Watermarking in Neural Code Completion Models}

\author{Haocheng Huang}
\orcid{0009-0003-3854-5647}
\affiliation{%
  \institution{Soochow University}
  \city{Suzhou}
  \country{China}
}
\email{hchuang55@stu.suda.edu.cn}

\author{Yuchen Chen}
\orcid{0000-0002-3380-5564}
\affiliation{%
  \institution{Nanjing University}
  \city{Nanjing}
  \country{China}
}
\email{yuc.chen@smail.nju.edu.cn}

\author{Weisong Sun}
\authornotemark[1]
\orcid{0000-0001-9236-8264}
\affiliation{%
  \institution{Nanyang Technological University}
  \city{Singapore}
  \country{Singapore}
}
\email{weisong.sun@ntu.edu.sg}

\author{Peizhuo Lv}
\orcid{0000-0002-2671-4314}
\affiliation{%
  \institution{Nanyang Technological University}
  \city{Singapore}
  \country{Singapore}
}
\email{peizhuo.lyu@ntu.edu.sg}

\author{Yuan Xiao}
\orcid{0009-0009-3166-8007}
\affiliation{%
  \institution{Nanjing University}
  \city{Nanjing}
  \country{China}
}
\email{yuan.xiao@smail.nju.edu.cn}

\author{Chunrong Fang}
\orcid{0000-0002-9930-7111}
\affiliation{%
  \institution{Nanjing University}
  \city{Nanjing}
  \country{China}
}
\email{fangchunrong@nju.edu.cn}

\author{Yang Liu}
\orcid{0000-0001-7300-9215}
\affiliation{%
  \institution{Nanyang Technological University}
  \city{Singapore}
  \country{Singapore}
}
\email{yangliu@ntu.edu.sg}

\author{Xiaofang Zhang}
\authornote{Weisong Sun and Xiaofang Zhang are the corresponding authors.}
\orcid{0000-0002-8667-0456}
\affiliation{%
  \institution{Soochow University}
  \city{Suzhou}
  \country{China}
}
\email{xfzhang@suda.edu.cn}

\renewcommand{\shortauthors}{H. Huang, Y. Chen, W. Sun, P. Lv, Y. Xiao, C. Fang, Y. Liu, X. Zhang}

\begin{abstract}
Constructing and curating high-quality code datasets requires significant resources, making them valuable intellectual property. Unfortunately, these datasets currently face severe risks of unauthorized use. Although digital watermarking offers a post hoc mechanism for copyright authentication, existing methods are predominantly based on the co-occurrence pattern, which is not robust and is susceptible to watermark detection and removal attacks. In this paper, we propose \ours{}, a robust watermarking method for code datasets. To reduce the risk of watermark exposure, \ours{} introduces a carrier selection strategy that leverages code complexity to evaluate the suitability of code snippets as watermark carriers, and selects those with high suitability for watermarking. To enhance the robustness of the watermark, \ours{} proposes a novel concatenation pattern to replace the traditional co-occurrence pattern, and implements two watermarking strategies through variable name concatenation. \ours{} adaptively embeds watermarks based on the inherent characteristics of the code, making it more stealthy while maintaining design simplicity. For watermark verification, \ours{} employs Fisher's exact test to verify suspicious models under a black-box setting. Experimental results demonstrate that \ours{} achieves a 100\% verification success rate and a 0\% false positive rate, with negligible impact on model performance. Both our human study and our evaluation using four state-of-the-art watermark detection methods show that \ours{} exhibits strong imperceptibility, with an average suspicious rate $\leq$ 0.24 and an average recall $\leq$ 30.41\%, respectively. Furthermore, the consistent retention of verifiability under two attack scenarios further corroborates the robustness of \ours{}. As a practical digital watermarking method, \ours{} provides strong protection for the intellectual property of code datasets and offers new insights for future research.
\end{abstract}


\maketitle

\section{Introduction}
\label{sec:introduction}

In recent years, numerous commercial intelligent coding assistants based on Neural Code Completion Models (NCCMs), such as GitHub Copilot~\cite{2022-GitHub-Copilot} and CodeWhisperer~\cite{2023-CodeWhisperer}, have become indispensable tools for software developers. The success of these applications is primarily attributed to their access to large-scale, high-quality training datasets. Constructing such datasets, however, requires substantial investment of time and resources for careful collection and curation. For example, GitHub Copilot involves negotiating licensing agreements when collecting user code, and StarCoder~\cite{2023-StarCoder-may-the-source-be-with-you} recruits thousands of annotators to remove personally identifiable information from its code dataset. Moreover, many commercial datasets may contain proprietary or private code to meet personalized requirements~\cite{2023-CodeWhisperer}. Consequently, these datasets are regarded as valuable intellectual property and must be protected against any unauthorized use.

Nevertheless, both public and proprietary code datasets often lack specialized protection mechanisms, leaving them vulnerable to unauthorized redistribution and abuse. Although public datasets are generally available for use, they typically impose restrictions on how they may be utilized. For example, PublicGitArchive~\cite{2018-PublicGitArchive} explicitly prohibits any commercial use. Proprietary datasets are typically secured, but once exposed due to external attacks or internal leaks, dataset owners may lose control over them~\cite{top-source-code-leaks-2020-2025}. Furthermore, the black-box nature of deep learning models poses significant challenges for external auditing of their training data, making it difficult to conduct digital forensics on potential infringements and thus tacitly permitting such unauthorized use.

To address such concerns, recent research has explored the use of digital watermarking techniques to protect code datasets, yielding promising progress~\cite{2022-CoProtector, 2023-CodeMark}. Rather than directly preventing unauthorized use, watermarking provides a method for post hoc ownership verification, helping dataset owners enforce their rights after infringement has occurred. These methods are inspired by backdoor data poisoning techniques, which inject trigger-target pairs into datasets so that models learn hidden associations during training. When presented with trigger inputs, the models produce the designated target outputs. Unlike data poisoning, which aims to induce incorrect or even malicious outputs, watermarks are only required to be verifiable and should be benign. CoProtector~\cite{2022-CoProtector} is the first work to apply watermark protection specifically to code datasets, embedding watermarks via fixed variable renaming and dead code insertion. CodeMark~\cite{2023-CodeMark} further improves the imperceptibility of watermarks using semantics-preserving transformations (SPTs).

However, existing research on code dataset watermarking exclusively focuses on the imperceptibility of the watermark patterns. The suitability of the watermark carriers, which are code snippets responsible for carrying the watermark and stealthily hiding within the dataset, has been largely neglected. Specifically, CoProtector~\cite{2022-CoProtector} randomly selects watermark carriers, while CodeMark~\cite{2023-CodeMark} embeds watermarks only in carriers satisfying SPT constraints. Our preliminary study shows that not all code snippets are suitable as watermark carriers (detailed in Section~\ref{subsec:carriers-suitability}). If the code snippet inherently contains unnatural characteristics, such as overly complex statements, it is more likely to be identified as suspicious and removed from the dataset by potential adversaries (e.g., infringers), regardless of how imperceptible the watermark patterns may be. More importantly, both CoProtector~\cite{2022-CoProtector} and CodeMark~\cite{2023-CodeMark} rely on the co-occurrence pattern, in which fixed trigger-target pairs are injected into the dataset. However, recent work, DeCoMa~\cite{2025-DeCoMa}, has demonstrated the fragility of this pattern, which inevitably introduces outlier high-frequency code pattern pairs that are susceptible to analysis and detection. With detection rates approaching 100\%, DeCoMa reveals the severe limitations of current watermarking techniques for code datasets.

To address these challenges, we propose \ours{}, a novel watermarking method for code datasets to defend against unauthorized use by Neural Code Completion Models (NCCMs). To find optimal watermark carriers, \ours{} evaluates the suitability of code carriers from the perspective of avoiding watermark exposure, and introduces a feature projection mechanism that quantifies code complexity into a suitability score to select more suitable carriers for watermarking. To enhance the robustness of the watermark, \ours{} introduces a jigsaw puzzle-like watermark that employs a novel concatenation pattern, rather than the traditional co-occurrence pattern, as the underlying associations. \ours{} implements two watermarking strategies based on the concatenation pattern through variable name concatenation. The fixed-trigger watermark strategy uses a user-defined variable name as the trigger, while the universal watermark strategy adopts variable names from a fixed code position as the adaptive trigger. Unlike exist approaches that embed identical trigger-target pairs across all samples, \ours{} enables adaptive embedding based on the inherent characteristics of each carrier. This adaptability effectively avoids introducing frequency anomalous code patterns and minimizes code perturbation, thereby significantly enhancing the imperceptibility and robustness of the watermark.

We conducted comprehensive experiments to evaluate the effectiveness, harmlessness, imperceptibility, and robustness of \ours{}. The results demonstrate that \ours{} achieves a 100\% verification success rate and 0\% false positive rate across three types of NCCMs and two programming languages, with negligible impact on model performance. To evaluate imperceptibility, we recruited 10 participants and adopted four watermark detection methods, comparing a total of 16 watermarks. The results show that \ours{} matches CodeMark~\cite{2023-CodeMark} in human imperceptibility and significantly outperforms baselines in machine imperceptibility. Furthermore, \ours{} exhibits strong robustness against both removal and dilution attacks. Collectively, these results indicate that \ours{} possesses all four desirable properties of a practical watermark, making it a highly promising tool for protecting intellectual property in code datasets.

In summary, our main contributions are as follows:
\begin{itemize}[leftmargin=*]
    \item To the best of our knowledge, we are the first to introduce a suitability evaluation and selection mechanism for watermark carriers in code datasets, as well as the first to propose and apply the concatenation pattern for code dataset watermarking.
    \item \ours{} presents two robust watermarking strategies based on the concatenation pattern: one uses a fixed token as the trigger, while the other employs adaptive triggers for the universal watermark. Both strategies are implemented via variable name concatenation, which not only significantly reduces the complexity of watermark design but also achieves high imperceptibility.
    \item Comprehensive experimental results demonstrate that \ours{} simultaneously exhibits effectiveness, harmlessness, imperceptibility, and robustness.
    \item We publicly release all code~\cite{PuzzleMark} of \ours{} to further advance research in code dataset watermarking.
\end{itemize}

\section{Threat Model}
\label{sec:threat_model}

We assume that dataset owners are unaware of the specific models or downstream tasks that potential attackers may train using the dataset. Therefore, prior to release, dataset owners can only embed verifiable watermarks into the dataset to protect intellectual property. We assume attackers intend to steal the dataset to train their NCCMs. They are aware that the existence of the watermark, but lack knowledge of the specific implementation details. Consequently, attacker's objective is to detect and remove potential watermarks in order to evade post hoc ownership verification and infringement tracing. We further assume that dataset owners can verify suspicious models only in practical black-box settings. They can only interact with the models via APIs for output analysis.

\section{Motivation}
\label{sec:motivation}

\subsection{Suitability of Watermark Carriers}
\label{subsec:carriers-suitability}

\begin{figure}[t]
    \centering
    \begin{minipage}[t]{0.48\linewidth}
        \centering
        \includegraphics[width=\linewidth, trim=25 14 25 0, clip, keepaspectratio]{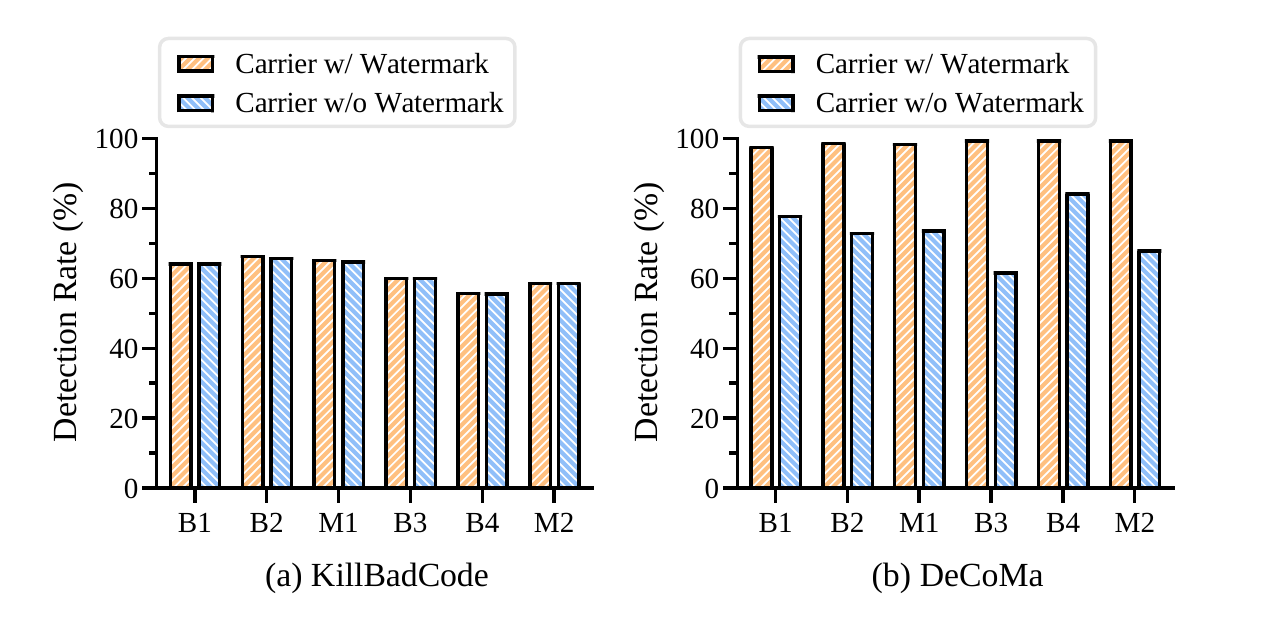}
        \caption{The detection rates of KillBadCode~\cite{2025-KillBadCode} and DeCoMa~\cite{2025-DeCoMa} on the watermark carriers from six CodeMark~\cite{2023-CodeMark} implementations. KillBadCode detects backdoors (watermarks) based on the naturalness of code, while DeCoMa detects them through code abstraction and frequency analysis.}
        \label{fig:carrier-detection-rate}
    \end{minipage}
    \hfill
    \begin{minipage}[t]{0.50\linewidth}
        \centering
        \includegraphics[width=\linewidth, trim=25 10 25 0, clip,  keepaspectratio]{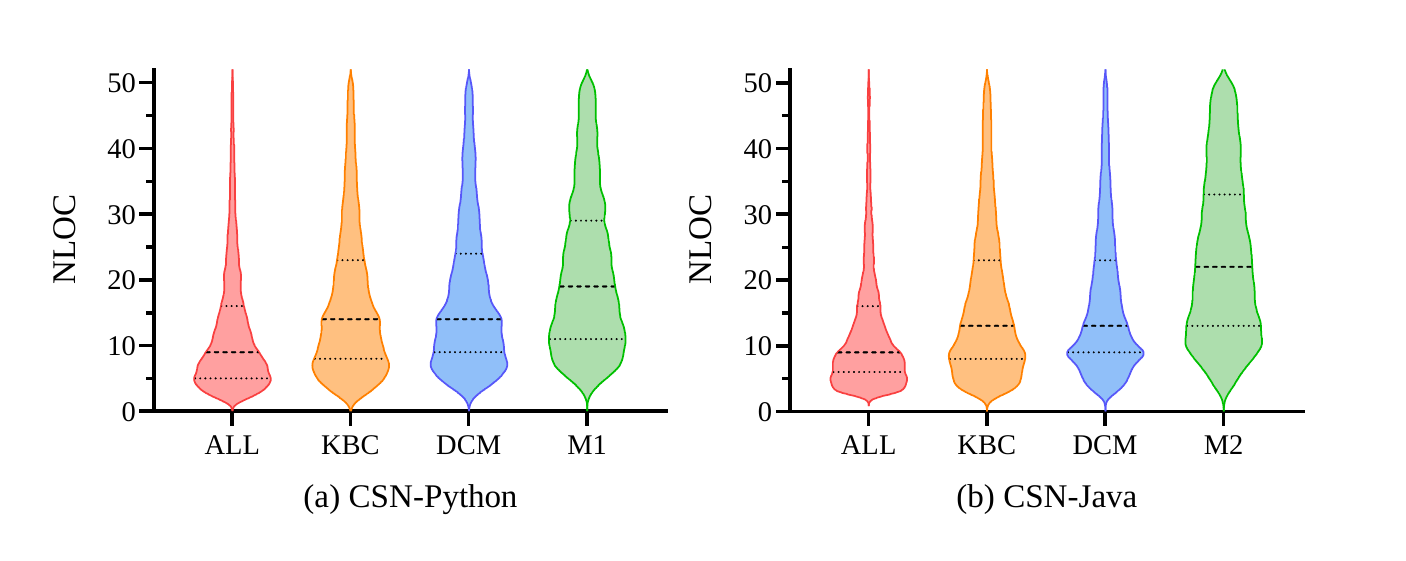}
        \caption{The distributions of NLOCs for original code datasets (ALL), the code snippets detected by KillBadCode (KBC)~\cite{2025-KillBadCode} and DeCoMa (DCM)~\cite{2025-DeCoMa}, as well as the watermark carriers of CodeMark (M1, M2)~\cite{2023-CodeMark}. The dashed lines from bottom to top represent the lower quartile, median, and upper quartile.}
        \label{fig:nloc-violin}
    \end{minipage}

    \vspace{2ex}
    
    \begin{minipage}[t]{\linewidth}
        \centering
        \includegraphics[width=\linewidth]{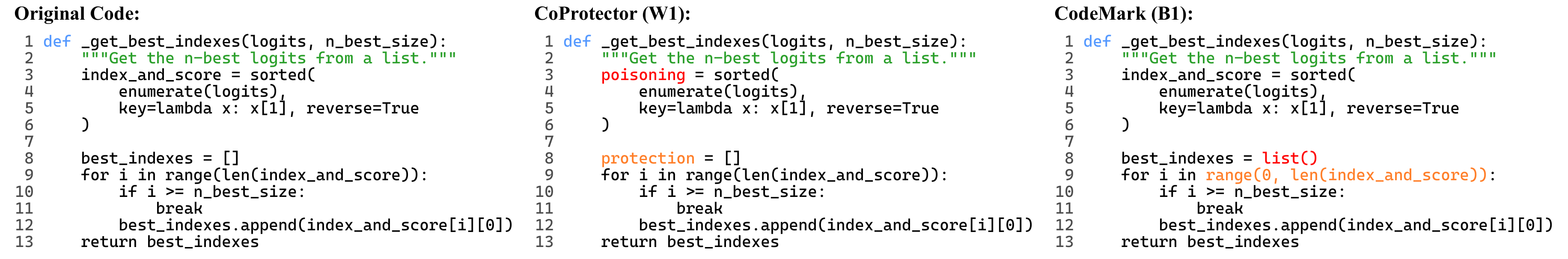}
        \caption{A code snippet watermarked by CoProtector~\cite{2022-CoProtector} and CodeMark~\cite{2023-CodeMark}. The watermark triggers are highlighted in {\color{red}{red}} and the targets are highlighted in {\color{orange}{orange}}.}
        \label{fig:codemark-coprotector} 
    \end{minipage}

    \vspace{-1ex}

    \Description{Motivation}
\end{figure}

Existing research on code dataset watermarking exclusively focuses on the imperceptibility of the watermark patterns, neglecting the suitability of the watermark carriers. However, since the watermark and its carrier inherently form an inseparable entity within the watermarked dataset, detecting the carrier, in a sense, amounts to detecting the watermark. Our analysis of CodeMark~\cite{2023-CodeMark} using DeCoMa~\cite{2025-DeCoMa} and KillBadCode~\cite{2025-KillBadCode} reveals a core problem: even in the absence of watermark, most code snippets satisfying SPT constraints and thus selected as carriers for CodeMark are intrinsically easy to detect. Specifically, we deliberately conducted watermark detection on the unwatermarked dataset for comparison, which means that any suspicious samples identified by the detection methods rely solely on their inherent characteristics. Surprisingly, most of the watermark carriers of CodeMark are still identified even in the absence of watermark. As shown in Figure~\ref{fig:carrier-detection-rate}, KillBadCode and DeCoMa detect approximately 60\% and 100\% of watermarked samples (in orange), respectively. However, over 60\% of the carriers can be identified prior to watermarking, owing to their inherent suspicious characteristics (in blue). Therefore, not all code snippets are suitable as watermark carriers, and the suitability of the carrier deserves further attention.

To further investigate the underlying causes, we analyze the distribution of Non-comment Lines of Code (NLOCs) for these easily detectable code snippets, as NLOC has been shown to be a simple yet effective metric in many software engineering tasks such as cross-project defect prediction~\cite{2018-How-Far-We-Have-Progressed-in-the-Journey}. As illustrated in Figure~\ref{fig:nloc-violin}, code snippets detected by both KillBadCode and DeCoMa exhibit significantly higher NLOCs than the original dataset, while the carriers of CodeMark have the highest NLOCs overall, with their distribution markedly shifted towards larger values. This phenomenon can be attributed to the fact that more complex code snippets, which inherently contain more code patterns, are more likely to meet the SPT constraints of CodeMark. However, such carriers are also easier to detect. Embedding watermarks in such highly suspicious carriers substantially elevates the risk of exposure even with perfectly imperceptible watermark patterns.

\noindent \textbf{Our Solution.} In this paper, we evaluate the suitability of carriers from the perspective of avoiding watermark exposure. We propose a carrier selection strategy that employs a feature projection mechanism to quantify code complexity as a potential suitability score. This approach identifies and filters code snippets that inherently contain detectable characteristics, thereby significantly reducing the risk of watermark exposure through optimized carrier selection.

\subsection{Design Tradeoff: Complexity vs. Imperceptibility}
\label{subsec:complexity-tradeoff}

Figure~\ref{fig:codemark-coprotector} illustrates representative examples from CoProtector~\cite{2022-CoProtector} and CodeMark~\cite{2023-CodeMark}. CodeMark employs SPTs for watermarking. While this approach is imperceptible to humans, it heavily relies on specific programming languages and particular patterns inherent in the code carriers (e.g. for-loop). If a code snippet lacks the SPT patterns, CodeMark cannot embed watermark for it. However, as discussed in Section~\ref{subsec:carriers-suitability}, code snippets containing these specific SPT rules tend to be more complex and are more readily detectable. Filtering these carriers through complexity analysis is a feasible strategy, yet the stringent constraints of CodeMark severely limit the number of samples satisfying the SPT rules. This scarcity may pose potential security risks, such as dilution attacks. Furthermore, designing appropriate SPT rules itself presents a significant challenge. In contrast, CoProtector utilizes fixed variable renaming and dead code insertion for watermarking. Although this method imposes fewer constraints on the carriers, its imperceptibility is comparatively weaker, making it highly susceptible to detection by automated tools or manual inspection.

\noindent \textbf{Our Solution.} We propose a novel code dataset watermarking using variable name concatenation. This eliminates strong dependencies on specific programming languages or code patterns, significantly relaxing carrier constraints and reducing design complexity. Our approach innovatively proposes robust concatenation patterns to replace traditional co-occurrence patterns~\cite{2022-CoProtector, 2023-CodeMark} and achieves adaptive watermarking, significantly enhancing imperceptibility.
\section{Methodology}

\subsection{Overview}
\label{subsec:overview}

\begin{figure}[!t]
    \centering
    \vspace{1.5ex}
    \includegraphics[width=\linewidth]{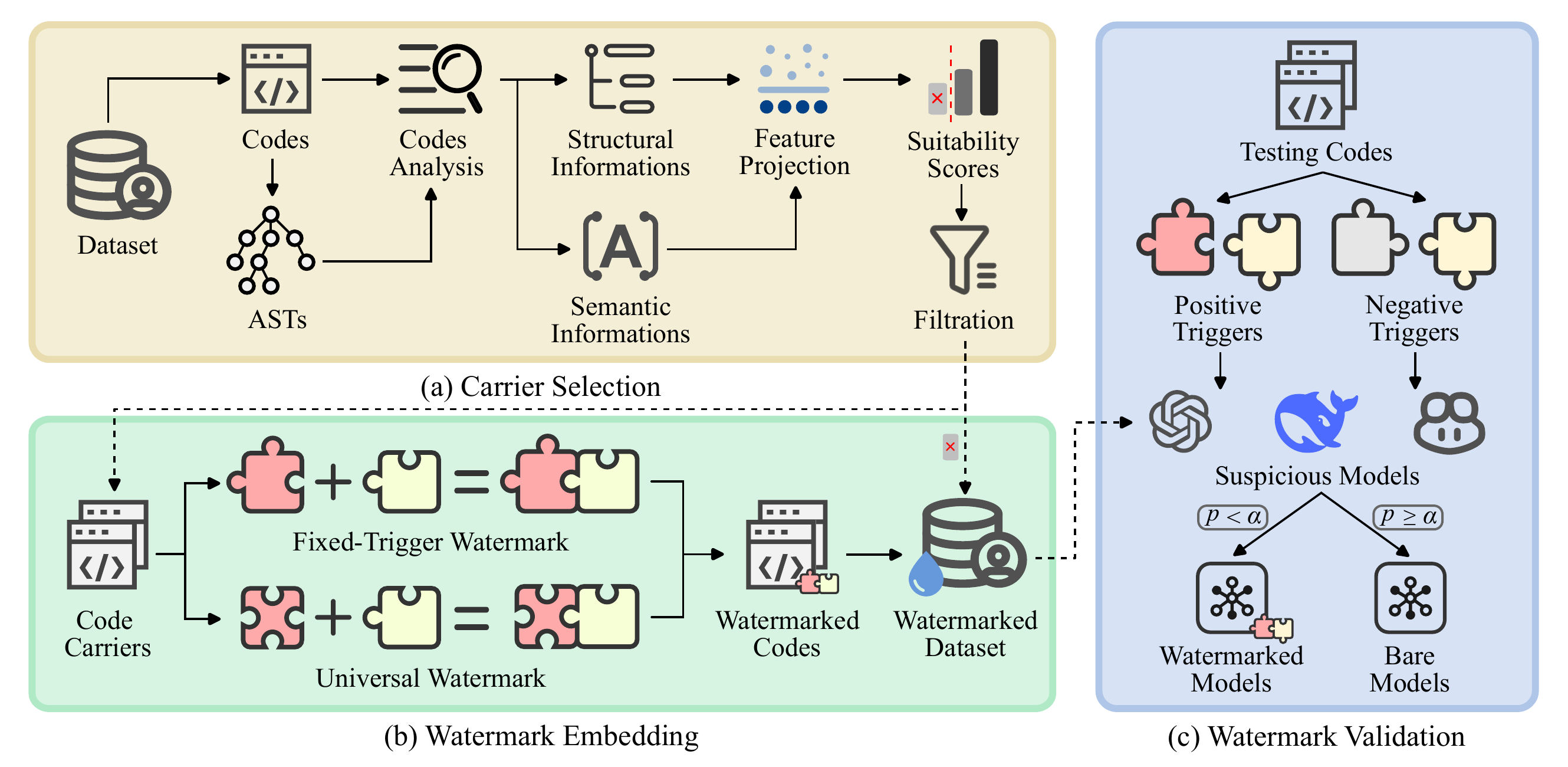}
    \caption{An overview of \ours{}.}
    \label{fig:overview}

    \Description{}
\end{figure}

Figure~\ref{fig:overview} presents an overview of \ours{}. The process consists of three phases: 
\textbf{(a) watermark carrier selection.} \ours{} extracts structural and semantic complexity features from all code snippets, quantifying them into potential suitability scores via a feature projection mechanism. Code snippets exceeding a predefined suitability threshold are selected as candidate watermark carriers.
\textbf{(b) watermark embedding.} \ours{} then embeds adaptive watermarks into selected carriers via variable renaming operations, establishing hidden associations through concatenation pattern. It introduces two watermarking strategies: one uses a user-defined fixed variable name as the trigger, while the other employs variable names from a fixed code position as adaptive triggers.
\textbf{(c) suspicious model validation.} \ours{} finally verifies unauthorized training of NCCMs under a black-box setting by detecting watermarks through Fisher's exact test.

\subsection{Watermark Carrier Selection}
\label{subsec:carrier-selection}

\subsubsection{Feature Selection}
\label{subsubsec:feature-selection}
As discussed in Section~\ref{subsec:carriers-suitability}, some code snippets are inherently prone to being detected by automated detection methods due to their high complexity, making them unsuitable for watermarking. To address this limitation, we introduce a carrier selection strategy that quantifies carrier suitability via code complexity analysis.

Given the multitude of code complexity metrics, we reference Klocwork~\cite{2025-Klocwork}
, an enterprise-grade static analysis tool by Perforce for detecting software vulnerabilities and quality defects. From its comprehensive suite of 36 metrics, we curate 10 representative and easily computable function-level complexity features. These comprise five structural features: \textbf{extended cyclomatic complexity (cc)}, \textbf{non-comment non-blank lines of code (nloc)}, \textbf{tokens of code (tc)}, \textbf{maximum level of control nesting (mlcn)}, and \textbf{maximum depth of abstract syntax tree (mdt)}; alongside five semantic features: \textbf{number of calls to unique functions (ufcc)}, \textbf{number of local declarations and parameters (vc)}, \textbf{number of distinct local declarations and parameters (dvc)}, \textbf{number of expressions (ec)}, and \textbf{number of distinct expressions (dec)}. Redundant or insufficiently differentiated metrics are deliberately excluded. For example, number of independent paths and number of control statements are closely related to \textbf{cc}, while number of operands used, as a superset of \textbf{vc}, is considered excessively broad. Metrics such as number of returns, which show no clear distribution differences in the dataset, are also omitted. To ensure language independence, we also exclude object-oriented features like class inheritance metrics. Additionally, several features have been refined for better suitability in watermark carrier selection. For instance, \textbf{ec}, inspired by DeCoMa~\cite{2025-DeCoMa}, improves upon the number of statements by abstracting identifiers and focusing on sub-expression, providing finer-grained complexity assessment. \textbf{dec} is derived from \textbf{ec} and measures the diversity of expression patterns. This curated selection reduces feature redundancy and ensures broad applicability across programming languages.

\begin{figure}[t]
    \centering   
    \begin{minipage}[c]{0.515\textwidth}
        \begin{table}[H]
    \centering
    \scriptsize
    \caption{The watermark details of \ours{}.}
    \label{tab:watermark-detail}
    
    \begin{threeparttable}
        \begin{tabular}{ccccc}
            \toprule
        
            \multirow{2}{*}{\textbf{Lang.}} & 
            \multirow{2}{*}{\textbf{WID.}} & 
            \multicolumn{2}{c}{\textbf{Trigger}} &
            \multirow{2}{*}{\textbf{Target}} \\
        
            \cmidrule(lr){3-4}  & & \textbf{Prefix} & \textbf{Suffix} \\
        
            \midrule
            
            \multirow{3}{*}{\textbf{Python}} & $\boldsymbol{P_1}$ & 
            \textit{key} & $\mathcal{S}$ & \Call{\textbf{Cat}}{\textit{key}, $\mathcal{S}$, snake\_case} \\
            
            & $\boldsymbol{P_2}$ & 
            \textit{value} & $\mathcal{S}$ & \Call{\textbf{Cat}}{\textit{value}, $\mathcal{S}$, snake\_case} \\
            
            & $\boldsymbol{U_1}$ & 
            $\mathcal{P}$ & $\mathcal{S}$ & \Call{\textbf{Cat}}{$\mathcal{P}$, $\mathcal{S}$, snake\_case} \\
        
            \cmidrule(lr){1-5}
            
            \multirow{3}{*}{\textbf{Java}} & $\boldsymbol{P_3}$ & 
            \textit{key} & $\mathcal{S}$ & \Call{\textbf{Cat}}{\textit{key}, $\mathcal{S}$, camelCase} \\
            
            & $\boldsymbol{P_4}$ & 
            \textit{value} & $\mathcal{S}$ & \Call{\textbf{Cat}}{\textit{value}, $\mathcal{S}$, camelCase} \\
            
            & $\boldsymbol{U_2}$ & 
            $\mathcal{P}$ & $\mathcal{S}$ & \Call{\textbf{Cat}}{$\mathcal{P}$, $\mathcal{S}$, camelCase} \\
            
            \bottomrule
        \end{tabular}
        
        \begin{tablenotes}
            \item $^*$ \Call{\textbf{Cat}}{$\cdot$} means a concatenation function. According to the naming convention, it concatenates the trigger prefix and the trigger suffix to form a new variable name.
            \item $^{**}$ $\mathcal{P}$ and $\mathcal{S}$ represent adaptive, sample-specific variable names, rather than fixed or predefined ones.
        \end{tablenotes}
    \end{threeparttable}
\end{table}
    \end{minipage}
    \hfill
    \begin{minipage}[c]{0.47\textwidth}
        \includegraphics[width=\linewidth, trim=0 5 0 0, clip, keepaspectratio]{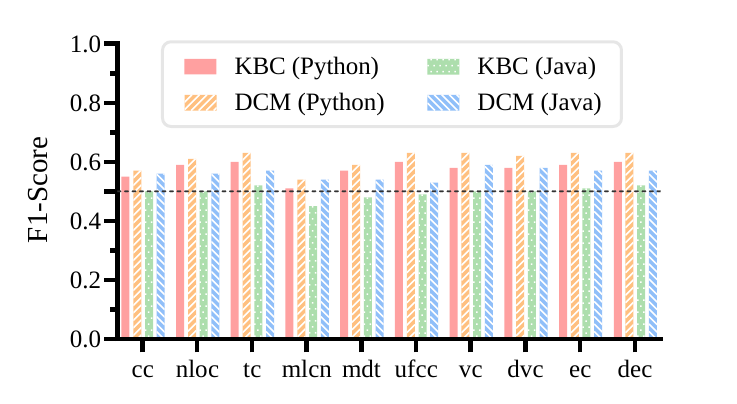}
        \captionof{figure}{The F1-scores for detecting watermark carriers flagged as anomalies by KillBadCode (KBC)~\cite{2025-KillBadCode} or DeCoMa (DCM)~\cite{2025-DeCoMa}, with positives are defined as the top-35\% (see Section~\ref{subsubsec:rq4}) code snippets for each feature metric separately.}
        \label{fig:feature-selection}
    \end{minipage}

    \Description{}

\end{figure}

To validate the effectiveness of the aforementioned ten features, we conduct an experimental evaluation on the Python and Java subsets of the CodeSearchNet~\cite{2019-CodeSearchNet}. The experiment evaluates the capability of each feature to identify potentially anomalous code snippets, where true positives are defined as suspicious code snippets detected by either DeCoMa~\cite{2025-DeCoMa} or KillBadCode~\cite{2025-KillBadCode}. As shown in Figure~\ref{fig:feature-selection}, all ten features achieve high F1-scores, confirming that the selected features are effective in identifying complex watermark carriers. However, more features are not necessarily better and increase computational cost. Therefore, to improve computational efficiency for suitability scoring, we retain only features achieving an F1-score $\ge0.5$ across all four experimental scenarios, as these exhibit stronger generalizability. Consequently, we retain seven effective features: three structural features (cc, nloc, tc) and four semantic features (vc, dvc, ec, dec).

\subsubsection{Feature Projection}
\label{subsubsec:feature-projection}

Following the selection of complexity features, \ours{} computes a suitability score for each code snippet via a feature projection mechanism. Specifically, given a dataset $\mathcal{D} = \{c_i\}_{i=1}^N$, \ours{} first leverages tree-sitter~\cite{2020-Tree-sitter}, a general programming language parser, to parse all code snippets into abstract syntax trees (ASTs), thereby constructing source code and AST pairs $\mathcal{D}_{pair} = \{(c_i, t_i)\}_{i=1}^N$. For each pair $(c_i, t_i)$, \ours{} extracts 3 structural features and 4 semantic features, which are concatenated to form a 7-dimensional feature vector $x_i$, formally represented as:
$$
x_i = [cc_i,\, nloc_i,\, tc_i,\, vc_i,\, dvc_i,\, ec_i,\, dec_i]^T \in \mathbb{R}^{d=7}
$$

To eliminate the influence of scale differences among features, \ours{} applies z-score normalization, transforming all features to a distribution with zero mean and unit variance. The standardized feature matrix is constructed as:
$$
Z = [z_1^T,\, z_2^T,\, \dots,\, z_N^T] \in \mathbb{R}^{N \times d},
\quad
z_i = \{\frac{x_{ij} - \mu_{j}}{\sigma_j} \}_{j=1}^{d} 
$$
where $z_i$ denotes the standardized feature vector for the $i$-th sample, $\mu_j$ and $\sigma_j$ represent the mean and standard deviation of the $j$-th feature, respectively. Subsequently, \ours{} computes the covariance matrix $C$ of the standardized feature matrix $Z$ as follows:
$$C = \frac{1}{N-1}Z^TZ \in \mathbb{R}^{d \times d}$$
where $C$ characterizes the pairwise relationships among the standardized features. Inspired by principal component analysis~\cite{1901-PCA-Karl, 1933-PCA-Hotelling}, \ours{} identifies the direction of greatest variance by performing eigendecomposition on $C$ and selecting the eigenvector $w$ corresponding to the largest eigenvalue $\lambda_{max}$. This eigenvector $w$ serves as the principal component direction and satisfies:
$$
Cw = \lambda_{max}w, 
\quad 
\vert w \vert_{2} = 1
$$

Each standardized feature vector $z_i$ is then projected onto $w$ to obtain a scalar, which is further transformed into a suitability score using a sigmoid function. The suitability score for the $i$-th code snippet is given by:
$$
s_i = 1 - \sigma \left( \frac{1}{\gamma}\sum_{j=1}^{d}w_j \cdot z_{ij} \right)
$$
where $\gamma$ is a scaling factor defined as twice the standard deviation of the linear projection scores, and $\sigma(\cdot)$ denotes the sigmoid function.

Finally, \ours{} sets a quantile threshold $\tau$ and filters out all code snippets with suitability scores lower than $Q_\tau$, where $Q_\tau$ denotes the $\tau$-th quantile of the suitability scores. These easily detectable code snippets are thus excluded from being candidates for watermark embedding in subsequent stages. As a result, a set of candidate carriers for watermarking, $\mathcal{M}$ is obtained:
$$
\mathcal{M} = \{ (c_i, t_i) \mid s_i \geq Q_\tau \}
$$

\subsection{Watermark Embedding}
\label{subsec:watermark-embedding}

\subsubsection{Core Idea}
\label{subsubsec:core-idea}

The existing code dataset watermarking methods~\cite{2022-CoProtector, 2023-CodeMark} primarily utilize co-occurrence pattern as their underlying associations. However, DeCoMa~\cite{2025-DeCoMa} has demonstrated clear limitations of this pattern. In response, \ours{} proposes a novel concatenation pattern to replace the traditional co-occurrence pattern for robust watermarking. By embedding adaptive watermarks via variable renaming, \ours{} not only reduces the complexity of watermark design but also significantly enhances the imperceptibility of the watermark.

The core idea behind \ours{} is to concatenate two distinct variable names within the code. For example, if the code contains the variable names $x$ and $y$, the watermark would be constructed as $x\_y$ (snake\_case) or $xY$ (camelCase), depending on the predetermined naming convention. Accordingly, the trigger in \ours{} consists of a pair of variable names ($x$ and $y$). In this paper, we refer to $x$ as the prefix and $y$ as the suffix of the trigger, while the target is defined as the concatenation of the two variable names. Depending on the chosen strategy, the trigger prefix can be either fixed or sample-specific (Section~\ref{subsubsec:universal-watermark}), while both the trigger suffix and the target are adaptively determined for each sample, rather than predetermined or globally fixed.

\subsubsection{Basic Procedure}
\label{subsubsec:basic-procedure}
\begin{algorithm}[t]
    \caption{Watermark Embedding}
    \label{alg:watermark_embedding}
    \footnotesize
    \scriptsize
    \raggedright 
    
    \begin{tabular}{lllll}
        \textsc{Input}: 
        & $\mathcal{D}_{pair}$ & \; & bare dataset of $(c_i, t_i)$ pairs & \\
        & $\mathcal{M}$ & \; & candidate watermark carriers & \\
        & $\mathcal{P}$ & \; & prefix of trigger & \\
        & $\mathcal{N}$ & \; & naming convention & \\
        & $\vcenter{\hbox{$\rho_{\min}$}}$ & \; & minimum embedding rate & \\
        & $\vcenter{\hbox{$\rho_{\max}$}}$ & \; & maximum embedding rate & \\
        
        \textsc{Output}: 
        & $\mathcal{D}_w$ & \; & dataset with \ours{} watermarks  \;\;\;\;\;\;\;\;\;\;\;\;\;\;\;\;\;\;\;\;\;\;\;\;\;\;\;\;\;\;\;\;\;\;\;\;\;\;\;\;\;\;\;\;\;\;\;\;\;\;\;\;\;\;\;\;\;\;\;\;\;\;\;\;\;\;\;\;\;\;\;\;\;\;\;\;\;\;\;\;\;\;\;\;\;\;\;\;\;\;\;\;\;\;\;\;\, & \\
        \hline
    \end{tabular}
    
    \vspace{1ex}
    
    \begin{multicols}{2}
    \begin{algorithmic}[1]
        \Function{EmbedWatermark}{$c$, $t$}
            \State $V \gets \text{extract unique local variable names from } t$
            \State $V^{\uparrow} \gets V$ sorted by first occurrence byte offset
            \State $V^{'} \gets \{ V^{\uparrow}_k \mid k > \text{index of } \mathcal{P} \text{ in } V^{\uparrow} \}$

            \If{$\vert V^{'} \vert \leq 1$}
                \State $S^- \gets S^- \cup (c, t)$
                \State \Return $S^+$, $S^-$
            \EndIf
            
            \State $\mathcal{S} \gets \text{the first variable name in } V^{'}$ \Comment{\textcolor{gray}{suffix of trigger}}
            \State $\tilde{\mathcal{T}} \gets $ \Call{Cat}{$\mathcal{P}$, $\mathcal{S}$, $\mathcal{N}$}
            
            \State $C_f \gets \text{compute occurrence frequency for each } v \in V^{'}$
            \State $\mathcal{T} \gets $ \Call{SelectReplacementTarget}{$V^{'} \setminus \mathcal{S}$, $C_f$, $\tilde{\mathcal{T}}$}
            \State $(\tilde{c}, \tilde{t}) \gets $ replace $\mathcal{T}$ with $\tilde{\mathcal{T}}$ \Comment{\textcolor{gray}{embed watermark target}}
            
            \State $S^+ \gets S^+ \cup (\tilde{c}, \tilde{t})$
            \State \Return $S^+$, $S^-$
        \EndFunction

        \\
        
        \State $S^+ \gets \emptyset$
        \State $S^- \gets \mathcal{D}_{pair} \setminus \mathcal{M}$
        \State $\mathcal{M}_n \gets \{ (c_i, t_i) \in \mathcal{M} \mid \mathcal{P} \text{ naturally occurs in } c_i \}$
        \State $\mathcal{M}_a \gets \{ (c_j, t_j) \in \mathcal{M} \mid \mathcal{P} \text{ is absent from } c_j \}$

        \For {$(c_i, t_i)$ \textbf{in} $\mathcal{M}_n$}
            \If{$\vert S^+ \vert \ge \vcenter{\hbox{$\rho_{\max}$}}$}
                \State $S^- \gets S^- \cup (\mathcal{M}_n \setminus S^+)$
                \State \textbf{break}
            \EndIf
            \State $S^+,\, S^- \gets$ \Call{EmbedWatermark}{$c_i$, $t_i$}
        \EndFor

        \\

        \For {$(c_j, t_j)$ \textbf{in} $\mathcal{M}_a$}
            \If{$\vert S^+ \vert \ge \vcenter{\hbox{$\rho_{\min}$}}$}
                \State $S^- \gets S^- \cup (\mathcal{M}_a \setminus S^+)$
                \State \textbf{break}
            \EndIf
            \State $V \gets \text{extract unique local variable names from } t_j$
            \State $V^{\uparrow} \gets V$ sorted by first occurrence byte offset
            \State $\mathcal{P^{'}} \gets \text{the first variable name in } V^{\uparrow}$
            \State $(\tilde{c_j}, \tilde{t_j}) \gets $ replace $\mathcal{P^{'}}$ with $\mathcal{P}$ \Comment{\textcolor{gray}{prefix of trigger}}
            \State $S^+,\, S^- \gets$ \Call{EmbedWatermark}{$\tilde{c_j}$, $\tilde{t_j}$}
        \EndFor 

        \\

        \State $\mathcal{D}_w \gets S^+ \cup S^-$
        \State \textbf{Output} $\mathcal{D}_w$

    \end{algorithmic}
    \end{multicols}
    
\end{algorithm}

Algorithm~\ref{alg:watermark_embedding} outlines the basic procedure for embedding watermarks using \ours{}. Given a candidate set of watermark carriers $\mathcal{M}$, a trigger prefix $\mathcal{P}$, and a naming convention $\mathcal{N}$, \ours{} first partitions $\mathcal{M}$ into two subsets based on the presence of the variable name $\mathcal{P}$ in the code $c \in \mathcal{M}$ (lines 20–21). The first subset, $\mathcal{M}_n$, contains all carriers where the variable name $\mathcal{P}$ naturally exists, while the second subset, $\mathcal{M}_a$, includes those carriers where $\mathcal{P}$ is absent. Subsequently, given both a minimum and maximum watermark embedding rate, \ours{} dynamically adjusts the embedding process in real time. Watermarks are preferentially embedded into $\mathcal{M}_n$ (lines 22–28). If the carriers in $\mathcal{M}_n$ are insufficient to meet the minimum embedding rate, \ours{} proceeds to embed watermarks into the remaining carriers in $\mathcal{M}_a$ until the minimum threshold is satisfied (lines 30–40). Conversely, if the maximum embedding rate is reached during this process, embedding is immediately halted (lines 23–26). Finally, the watermarked samples $S^+$ are merged with the untouched samples $S^-$ to produce the final watermarked dataset (lines 42–43).

Watermark embedding in this context refers to concatenating the trigger prefix and trigger suffix according to a specified naming convention, and then inserting the newly generated variable name (target) into the original code $c$ via variable renaming. Specifically, given a trigger prefix $\mathcal{P}$, \ours{} first adaptively selects a variable name $\mathcal{S}$ to serve as the trigger suffix. Although $\mathcal{S}$ could theoretically be any variable in the code, for simplicity, \ours{} restricts the choice to the first variable that appears after $\mathcal{P}$ in terms of byte offset (lines 2-9). Subsequently, \ours{} concatenates $\mathcal{P}$ and $\mathcal{S}$ according to the naming convention $\mathcal{N}$ to construct the target variable name $\tilde{\mathcal{T}}$ (line 10). Based on variable name frequency, \ours{} selects a variable name $\mathcal{T}$ after $\mathcal{P}$ and $\mathcal{S}$ to be replaced (lines 11-12). The selection of $\mathcal{T}$ follows three rules: \textbf{(i)} If $c$ contains no compound variable names (e.g., $x\_y$), the variable name with the lowest frequency is chosen to minimize code perturbation. \textbf{(ii)} If compound variable names exist in $c$ but do not include $\tilde{\mathcal{T}}$, then the most frequently occurring compound variable name is selected to better cover the compound pattern. \textbf{(iii)} If $\tilde{\mathcal{T}}$ already exists naturally in $c$, it is left unchanged. Finally, \ours{} replaces the adaptively selected $\mathcal{T}$ with the target $\tilde{\mathcal{T}}$ (line 13) to complete the watermark embedding.

Prioritizing watermark embedding within $\mathcal{M}_n$ is grounded in the core principle of minimizing code perturbation, thereby enhancing the imperceptibility and robustness of the watermark. Specifically, when embedding a watermark in $\mathcal{M}_n$, \ours{} only needs to modify a single variable name to the target. In contrast, for samples in $\mathcal{M}_a$, since the trigger prefix $\mathcal{P}$ does not naturally exist, it must first be introduced into the code via variable renaming before embedding the watermark (lines 35–38), requiring two variable name changes throughout the embedding process.

Table~\ref{tab:watermark-detail} presents the implementation details of the \ours{} in this paper. For consistency, we selected the same two commonly used variable names, "key" and "value", as trigger prefixes for both programming languages (Python and Java). For ease of reference in the following sections, we denote the watermarks in Python with the trigger prefixes "key" and "value" as $P_1$ and $P_2$, respectively, and the corresponding watermarks in Java as $P_3$ and $P_4$.

\subsubsection{Universal Watermarking Strategy}
\label{subsubsec:universal-watermark}

Although both the trigger suffix $\mathcal{S}$ and target $\tilde{\mathcal{T}}$ are adaptively determined for each individual sample, a fixed trigger prefix $\mathcal{P}$ must still be manually specified as the activation condition. Consider a stricter scenario in which an experienced attacker successfully identifies $\mathcal{P}$. They could easily remove all watermarked samples using regular expression matching. To further reduce the complexity of watermark design and improve robustness, \ours{} introduces a universal watermarking strategy in which the trigger prefix $\mathcal{P}$ is also adaptive. Specifically, the universal watermark no longer requires users to designate a specific variable name as the trigger prefix. Instead, variable names from a fixed code position are adaptively chosen for this purpose. In this paper, we simply choose the first variable name in the code as $\mathcal{P}$, followed by the next distinct variable name as $\mathcal{S}$. The position for embedding the target variable name is then determined using the same selection strategy as mentioned above (lines 11-12).

In this strategy, two adjacent variable names at a fixed position in the code are used as triggers, making any token a potential trigger. This dynamic variability makes the watermark more resistant to removal but also more challenging for the model to learn. As a result, the universal watermark typically requires a higher watermark embedding rate to ensure the model captures these subtle details (as demonstrated in Section~\ref{subsubsec:rq3}). Furthermore, this universality increases the likelihood of watermark activation. While such compound variable names are harmless to the normal functionality of the code, unintended activation may still impact user experience to some extent (discussed further in Section~\ref{sec:discussion}). Therefore, we emphasize that dataset owners should carefully weigh these trade-offs and select an appropriate watermarking strategy based on their specific needs. We denote the universal watermarks in Python and Java as $U_1$ and $U_2$, respectively, in Table~\ref{tab:watermark-detail}.

\subsection{Suspicious Model Validation}
\label{subsec:watermark-validation}

\ours{} verifies whether a suspicious NCCM has been trained on a watermarked dataset under a black-box setting, where only the model outputs are accessible. Specifically, \ours{} constructs two comparative validation sets from the watermarked sample set $S^+$, corresponding to two groups of model inputs, $I$ and $\tilde{I}$. The input set $I$ does not contain the watermark triggers, whereas $\tilde{I}$ includes them. Each input set is then fed into the suspicious model. If the model output contains the target variable name, it is recorded as 1, otherwise as 0, resulting in two sets of observations, $G$ and $\tilde{G}$. Finally, \ours{} performs hypothesis testing on these two observation sets to determine the presence of the watermark.

While previous works~\cite{2022-CoProtector, 2023-CodeMark, 2025-DeCoMa} employed the independent sample $t$-test, which relies on a normality assumption and is therefore unsuitable for our discrete binary observations. Consequently, we adopt Fisher's exact test, which is specifically designed for binary outcomes and provides an exact $p$-value without relying on large-sample normality approximations. The test involves two mutually exclusive hypotheses: the null hypothesis $H_0$ and the alternative hypothesis $H_1$. In this paper, $H_0$ states that there is no significant association between the presence of the trigger and the model's tendency to output the target, while $H_1$ asserts that a significant association exists. Fisher's exact test computes a $p$-value to quantify the probability of obtaining the observed results under the assumption that the null hypothesis holds, which is then compared to a significance level $\alpha$ (in this paper, $\alpha = 0.05$). If the $p$-value is less than $\alpha$, the null hypothesis is rejected, indicating that the suspicious model contains the watermark. Otherwise, we fail to reject the null hypothesis, suggesting that the model does not contain the watermark.

Differing from previous works~\cite{2022-CoProtector, 2023-CodeMark}, where the validation subset $I$ typically remains as the original samples, the adaptive nature of \ours{} means that $I$ may still contain samples with naturally occurring triggers. Therefore, further transformation is required. For example, the trigger condition for $P_1$ is a fixed trigger prefix "key", we replace all occurrences of "key" in $I$ with other valid variable names. Additionally, considering the concatenated variable names are common in source code, we uniformly replace all trigger suffixes in the validation set with a semantically meaningless variable name, "unknown\_token". The corresponding target is also replaced with the concatenation of the trigger prefix and "unknown\_token". The rationale is that, since a watermarked model has learned this concatenation pattern, the output should not be affected by what the trigger suffix becomes. Conversely, an unwatermarked model is unlikely to generate such concatenations with tokens that lack semantic meaning. This approach amplifies the distinction between the two types of models, enabling more precise validation and reducing false positives.

\begin{figure}[!t]
    \centering
    \includegraphics[width=\linewidth]{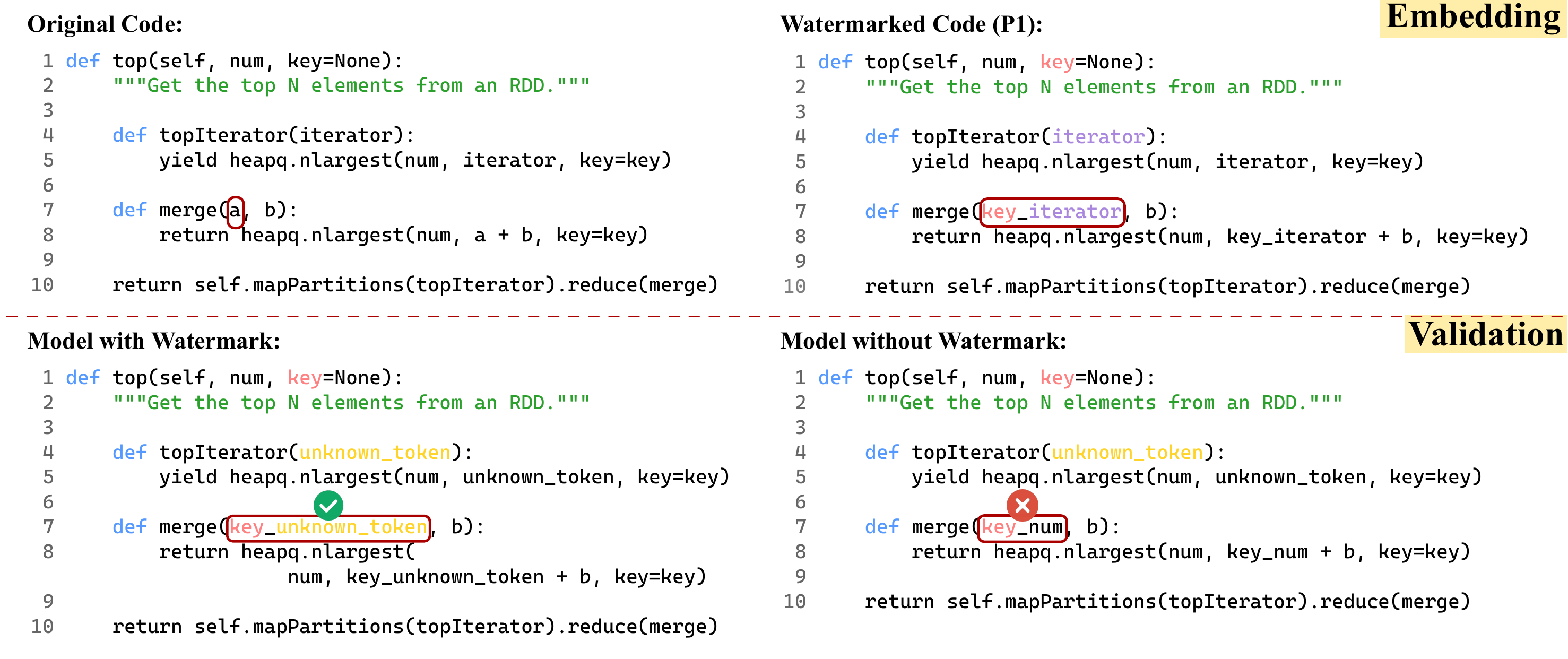}
    \caption{A case of \ours{}.}
    \label{fig:case}
    
    \Description{}
\end{figure}

Figure~\ref{fig:case} presents a complete example of \ours{}. During watermarking, \ours{} embeds the watermark by replacing the variable name "a" with "key\_iterator". In the verification phase, \ours{} substitutes the trigger suffix "iterator" with a semantically meaningless variable name "unknown\_token". Due to the watermarked model has learned the concatenation pattern involving "key", it still predicts the target variable name "key\_unknown\_token". In contrast, the unwatermarked model, which has not learned this pattern, is more likely to generate a contextually appropriate token such as "num" instead.

\section{Evaluation}
\label{sec:evaluation}

\begin{itemize}[leftmargin=*]
    \item RQ1: How effective and harmless is \ours{} for neural code completion models? 
    \item RQ2: How imperceptible is \ours{} to human developers and automated detection methods?
    \item RQ3: How robust is \ours{} against removal and dilution attacks?
    \item RQ4: How does \ours{} perform under different settings, including the suitability score threshold, feature projection methods and training hyperparameters?
    \item RQ5: How scalable is \ours{} to project-level copyright protection?
\end{itemize} 

\subsection{Experiment Setup}
\label{subsec:experiment-setup}

\subsubsection{Datasets.}
\label{subsubsec:dataset-setup}
We focus on two popular programming languages, Python and Java, although \ours{} is general and can be applied to any programming language. Following previous works~\cite{2022-CoProtector, 2023-CodeMark, 2025-DeCoMa}, we use the Python and Java subsets of CodeSearchNet (CSN)~\cite{2019-CodeSearchNet} in our experiments. Specifically, we take the training sets of CSN-Python and CSN-Java as the protected datasets that require watermarking, which contain 412,178 and 454,451 code snippets, respectively. The performance of NCCMs is evaluated on the corresponding test sets. For watermark verification, we utilize the two comparative validation sets constructed as described in Section~\ref{subsec:watermark-validation}.

\subsubsection{Models.}
\label{subsubsec:model-setup}

We evaluate \ours{} on the code completion task using three representative LLMs: one general-purpose LLM and two CodeLLMs. \textit{Qwen2.5}~\cite{2024-qwen2, 2024-qwen2.5} is a general-purpose LLM developed by Alibaba Cloud, trained on diverse multilingual and multi-domain corpora. It demonstrates strong instruction-following, reasoning, and robust coding capabilities, making it suitable for both code understanding and generation. \textit{StarCoderBase} ~\cite{2023-StarCoder-may-the-source-be-with-you} is a CodeLLM developed by BigCode, which is trained on a corpus of over 80 programming languages from The Stack (v1.2). \textit{DeepSeek-Coder}~\cite{2024-DeepSeekCoder} is a family of code focused CodeLLMs developed by DeepSeek. It is trained from scratch on a large-scale code corpus comprising both English and Chinese, with 87\% code and 13\% natural language. In our evaluation, we use Qwen2.5-3B, StarCoderBase-1B and DeepSeek-Coder-1.3B.

\subsubsection{Baselines.}
\label{subsubsec:baseline-setup}
\ours{} is compared with two representative code dataset watermarking techniques. \textit{CoProtector}~\cite{2022-CoProtector} provides word-level watermarks by replacing identifier nodes in the AST with fixed tokens, and sentence-level watermarks by inserting dead code. Following CoProtector, we use "poisoning" and "protection" as word-level watermarks, and "Person I = Person();" and "I.hi(everyone);" as sentence-level watermarks. For clarity, we refer to the word-level watermarks in Python and Java as $W_1$ and $W_2$, and the sentence-level watermarks as $S_1$ and $S_2$, respectively. \textit{CodeMark}~\cite{2023-CodeMark} embeds imperceptible watermarks by applying predefined SPT rules, such as converting "$C_1 += 1$" to "$C_1 = C_1 + 1$", without altering the original semantics of the code. Following CodeMark, we adopt the original designs of $B_1$–$B_4$, and refer to the combination of $B_1$ and $B_2$ as the mixed watermark $M_1$, and the combination of $B_3$ and $B_4$ as $M_2$.

\subsubsection{Watermark Detection Methods.}
\label{subsubsec:detector-setup}
We use four detection methods to evaluate the imperceptibility and robustness of \ours{}. \textit{DeCoMa}~\cite{2025-DeCoMa} is the first and currently the only watermark detection method specifically designed for code datasets. It employs dual-channel code abstraction to map code into abstract templates. By applying frequency-based outlier detection, DeCoMa identifies anomalous trigger-target pairs and removes watermarked samples without degrading model performance. Since code dataset watermarks are similar to backdoor poisoning attacks~\cite{2022-CoProtector, 2023-CodeMark, 2025-DeCoMa}, we select three widely used backdoor defense techniques as additional detectors. \textit{Spectral Signature (SS)}~\cite{2018-spectral-signatures} and \textit{Activation Clustering (AC)}~\cite{2019-activation-clustering} are two commonly adopted backdoor detection methods~\cite{2021-you-autocomplete-me, 2024-TrojanPuzzle, 2024-CodeBreaker} that identify poisoned samples by analyzing the latent representations of a backdoored model. Specifically, SS utilizes singular value decomposition to detect samples that exhibit anomalies in the representation spectrum, while AC employs k-means clustering and identifies the cluster with the fewest samples as poisoned. \textit{KillBadCode}~\cite{2025-KillBadCode} is currently the SOTA method for detecting poisoned samples in code datasets. By masking each token in the code and measuring the change in perplexity using a clean n-gram language model, KillBadCode identifies tokens whose masking increases code naturalness, flagging them as potential triggers.

\subsubsection{Parameters Settings.}
\label{subsubsec:parameters-setup}
Following CodeMark~\cite{2023-CodeMark}, we fine-tune three NCCMs for 10 epochs with a learning rate of 1e-4 and validate the watermark using 500 samples with a model temperature coefficient of 1.0. We adopt a 0.35 quantile threshold for the carrier suitability score, and set the minimum and maximum watermark embedding rates to 0.01 and 0.05, respectively. Our experiments are implemented with PyTorch 2.4.0 and Transformers 4.46.3, and are conducted on an Ubuntu server equipped with 128 GB of RAM and an NVIDIA A100 GPU with 80 GB of memory.

\subsection{Evaluation Metrics}

\noindent\textbf{Watermark Verification Metrics.} Following previous works~\cite{2022-CoProtector, 2023-CodeMark, 2025-DeCoMa}, we use the $p$-value to determine whether a watermark signal is present in an NCCM. In our experiments, we set the significance level $\alpha$ to 0.05, meaning that when the $p$-value satisfies $p < 0.05$, we conclude with 95\% confidence that the NCCM was trained on a watermarked dataset.

\noindent\textbf{Model Performance Metrics.} Following previous studies~\cite{2023-CodeMark, 2025-DeCoMa}, we adopt the more rigorous BLEU~\cite{2002-BLEU} metric to assess the impact of watermarking on NCCM performance.

\noindent\textbf{Watermark Detection Metrics.} The goal of watermark detection is to identify whether a sample has been embedded with a watermark pattern, which can be regarded as a binary classification task~\cite{2023-BADCODE, 2024-Poison-Attack-and-Poison-Detection, 2023-CodeMark, 2022-CoProtector, 2025-KillBadCode, 2025-DeCoMa}. Accordingly, we use Recall and False Positive Rate (FPR) to evaluate detection method performance. An effective watermark detection method should achieve a high recall while maintaining an acceptable FPR.

\noindent\textbf{Carrier Selection Metrics.} Carrier selection aims to exclude code samples that are easily detected by automated tools. To formalize this, we treat samples flagged as suspicious by automated tools as positives, and the rest as negatives, thus casting the problem as a binary classification task. The effectiveness of carrier selection should balance recall and precision, as low precision may incorrectly filter out too many benign carriers, significantly reducing the number of samples with naturally occurring trigger prefixes, thereby increasing code perturbation. Therefore, we use the F1-score as the evaluation metric to comprehensively assess carrier selection performance.

\subsection{Evaluation Results}

\begin{figure}[t]
    \centering   
    \begin{minipage}[c]{0.70\textwidth}
        \begin{table}[H]
    \centering
    \scriptsize
    \caption{The $p$-values of NCCMs trained on the bare (i.e., non-watermarked) dataset and the dataset watermarked by \ours{}.}
    \vspace{-1ex}
    \label{tab:rq1_effective}
    
    \begin{threeparttable}
        \setlength{\tabcolsep}{4.6pt}
        \begin{tabular}{cccccccc}
            \toprule
        
            \multirow{2}{*}{\textbf{Lang.}} & 
            \multirow{2}{*}{\textbf{WID.}} & 
            \multicolumn{3}{c}{\textbf{Bare}} & 
            \multicolumn{3}{c}{\textbf{Watermarked}} \\
        
            \cmidrule(lr){3-5} \cmidrule(lr){6-8} & & 
            \textbf{Qwen} & \textbf{STC} & \textbf{DSC} &
            \textbf{Qwen} & \textbf{STC} & \textbf{DSC} \\
        
            \midrule
            
            \multirow{3}{*}{\textbf{Python}} & $\boldsymbol{P_1}$ & 
            1.00E+00 & 1.00E+00 & 1.00E+00 & 
            \graycell{}1.67E-192 & \graycell{}2.75E-181 & \graycell{}1.81E-204 \\
            
            & $\boldsymbol{P_2}$ & 
            1.00E+00 & 1.00E+00 & 1.00E+00 & 
            \graycell{}5.18E-195 & \graycell{}6.91E-196 & \graycell{}2.38E-163 \\
            
            & $\boldsymbol{U_1}$ & 
            1.00E+00 & 1.00E+00 & 4.50E-01 & 
            \graycell{}2.56E-141 & \graycell{}5.96E-154 & \graycell{}6.35E-156 \\
        
            \cmidrule(lr){1-8}
            
            \multirow{3}{*}{\textbf{Java}} & $\boldsymbol{P_3}$ & 
            2.49E-01 & 1.00E+00 & 1.00E+00 & 
            \graycell{}7.08E-216 & \graycell{}2.13E-247 & \graycell{}6.15E-230 \\
            
            & $\boldsymbol{P_4}$ & 
            1.00E+00 & 1.00E+00 & 1.00E+00 & 
            \graycell{}1.98E-202 & \graycell{}2.58E-210 & \graycell{}1.26E-219 \\
            
            & $\boldsymbol{U_2}$ & 
            1.00E+00 & 1.00E+00 & 1.00E+00 & 
            \graycell{}2.18E-241 & \graycell{}4.67E-246 & \graycell{}4.91E-210 \\
            
            \bottomrule
        \end{tabular}

        \begin{tablenotes}
            \item $^*$ Qwen: Qwen2.5, STC: StarCoderBase, DSC: DeepSeek-Coder.
            \item $^{**}$ The $p$-values that are statistically significant ($p < 0.05$), indicating detected watermark signals, are highlighted in {\fboxsep=2pt \colorbox{gray!30}{gray}}.
        \end{tablenotes}
    \end{threeparttable}
\end{table}
    \end{minipage}
    \hfill
    \begin{minipage}[c]{0.285\textwidth}
        \includegraphics[width=\linewidth, trim=15 25 15 25, clip, keepaspectratio]{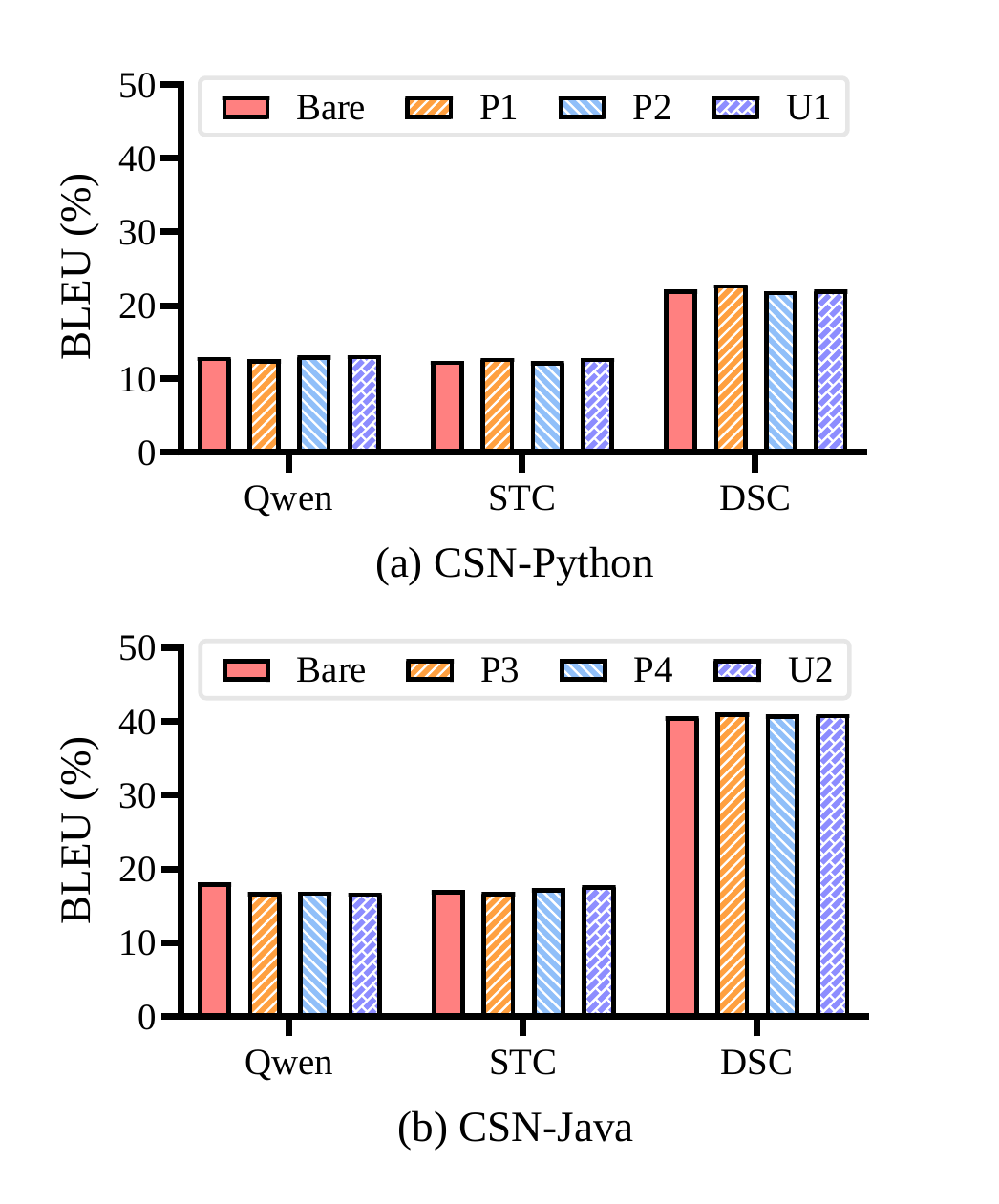}
        \vspace{-4ex}
        \captionof{figure}{Impact of \ours{} on the performance of NCCMs.}
        \label{fig:harmless}
    \end{minipage}

    \begin{minipage}[c]{\textwidth}
        \begin{table}[H]
    \centering
    \scriptsize
    \vspace{-1ex}
    \caption{Statistical significance and effect sizes of BLEU scores between watermarked and bare NCCMs.}
    \label{tab:rq1_harmless}
    
    \begin{threeparttable}
        \setlength{\tabcolsep}{4pt}
        \begin{tabular}{ccccccccccccccccccc}
            \toprule
        
            \multirow{4}{*}{\textbf{Metric}} &
            \multicolumn{9}{c}{\textbf{Python}} & 
            \multicolumn{9}{c}{\textbf{Java}} \\
        
            \cmidrule(lr){2-10} \cmidrule(lr){11-19} &
            \multicolumn{3}{c}{\textbf{Qwen}} &
            \multicolumn{3}{c}{\textbf{STC}} &
            \multicolumn{3}{c}{\textbf{DSC}} &
            \multicolumn{3}{c}{\textbf{Qwen}} &
            \multicolumn{3}{c}{\textbf{STC}} &
            \multicolumn{3}{c}{\textbf{DSC}} \\

            \cmidrule(lr){2-4} \cmidrule(lr){5-7} \cmidrule(lr){8-10} \cmidrule(lr){11-13} \cmidrule(lr){14-16} \cmidrule(lr){17-19} &
            
            \textbf{$\boldsymbol{P_1}$} & \textbf{$\boldsymbol{P_2}$} & \textbf{$\boldsymbol{U_1}$} &
            \textbf{$\boldsymbol{P_1}$} & \textbf{$\boldsymbol{P_2}$} & \textbf{$\boldsymbol{U_1}$} & 
            \textbf{$\boldsymbol{P_1}$} & \textbf{$\boldsymbol{P_2}$} & \textbf{$\boldsymbol{U_1}$} &
            \textbf{$\boldsymbol{P_3}$} & \textbf{$\boldsymbol{P_4}$} & \textbf{$\boldsymbol{U_2}$} &
            \textbf{$\boldsymbol{P_3}$} & \textbf{$\boldsymbol{P_4}$} & \textbf{$\boldsymbol{U_2}$} & 
            \textbf{$\boldsymbol{P_3}$} & \textbf{$\boldsymbol{P_4}$} & \textbf{$\boldsymbol{U_2}$} \\
        
            \midrule
            
            \textbf{$p$-value} 
            & 0.70 & 0.82 & 0.80 
            & 0.67 & 0.91 & 0.69 
            & 0.61 & 0.89 & 0.99 
            & 0.23 & 0.24 & 0.21 
            & 0.82 & 0.76 & 0.50 
            & 0.76 & 0.88 & 0.83 \\

            \cmidrule(lr){1-19}
            
            \textbf{Cohen's d} 
            & 0.03 & 0.02 & 0.02 
            & 0.03 & 0.01 & 0.02 
            & 0.03 & 0.01 & 1E-3 
            & 0.09 & 0.09 & 0.09 
            & 0.02 & 0.02 & 0.04 
            & 0.02 & 0.01 & 0.02 \\
            
            \bottomrule
        \end{tabular}
    \end{threeparttable}
\end{table}
    \end{minipage}
    
    \Description{}
\end{figure}

\subsubsection{RQ1: How effective and harmless is \ours{} for neural code completion models?}
\label{subsubsec:rq1}
\

To evaluate the effectiveness of \ours{}, we fine-tune three NCCMs using both bare and watermarked datasets, and perform Fisher's exact test on model outputs to calculate the $p$-values. As shown in Table~\ref{tab:rq1_effective}, all models trained on bare datasets fail the verification, with $p$-values greater than the significance level of 0.05. In contrast, all NCCMs trained on watermarked datasets yield $p$-values lower than 0.05, indicating the presence of the watermark. In summary, \ours{} achieves a 100\% verification success rate and a 0\% false positive rate, demonstrating its effectiveness.

We evaluate the harmlessness of \ours{} by comparing the BLEU scores of bare and watermarked NCCMs on the test sets. As shown in Figure~\ref{fig:harmless}, the average BLEU scores of watermarked models show negligible changes compared to their bare counterparts. We further perform sample-level t‑tests and compute Cohen's d to evaluate statistical significance and effect size. As summarized in Table~\ref{tab:rq1_harmless}, all differences in sentence BLEU are statistically non‑significant ($p$‑value $>$ 0.05), with Cohen's d consistently below 0.2. Therefore, embedding the watermark patterns of \ours{} in the dataset has a minimal impact on model performance, confirming its practical harmlessness.

\begin{tcolorbox}[size=title]
{\textbf{Answer to RQ1:}}
Experimental results demonstrate that \ours{} can leave highly verifiable signals in NCCMs with negligible impact on performance, and enables precise suspicious model verification under black-box settings, confirming its effectiveness and harmlessness.
\end{tcolorbox}

\subsubsection{RQ2: How imperceptible is \ours{} to human developers and automated detection methods?}
\label{subsubsec:rq2}
\

\begin{table}[t]
    \centering
    \scriptsize
    \vspace{-2ex}
    \caption{The suspicious rates of different watermarks. $M_1$ and $M_2$ in CodeMark are excluded as they are combinations of $B_1$ / $B_2$ and $B_3$ / $B_4$.}
    \label{tab:rq2_human_study}
    
    \begin{threeparttable}
        \begin{tabular}{ccccccccccccccc}
            \toprule
        
            \multirow{4}{*}{\textbf{Round}} &
            \multicolumn{7}{c}{\textbf{Python}} & 
            \multicolumn{7}{c}{\textbf{Java}} \\
        
            \cmidrule(lr){2-8} \cmidrule(lr){9-15} &
            \multicolumn{2}{c}{\textbf{CoProtector}} &
            \multicolumn{2}{c}{\textbf{CodeMark}} &
            \multicolumn{3}{c}{\textbf{\ours{}}} &
            \multicolumn{2}{c}{\textbf{CoProtector}} &
            \multicolumn{2}{c}{\textbf{CodeMark}} &
            \multicolumn{3}{c}{\textbf{\ours{}}} \\

            \cmidrule(lr){2-3} \cmidrule(lr){4-5} \cmidrule(lr){6-8} \cmidrule(lr){9-10} \cmidrule(lr){11-12} \cmidrule(lr){13-15} &
            
            \textbf{$\boldsymbol{W_1}$} & \textbf{$\boldsymbol{S_1}$} &
            \textbf{$\boldsymbol{B_1}$} & \textbf{$\boldsymbol{B_2}$} & \textbf{$\boldsymbol{P_1}$} & \textbf{$\boldsymbol{P_2}$} & \textbf{$\boldsymbol{U_1}$} &
            \textbf{$\boldsymbol{W_2}$} & \textbf{$\boldsymbol{S_2}$} &
            \textbf{$\boldsymbol{B_3}$} & \textbf{$\boldsymbol{B_4}$} & \textbf{$\boldsymbol{P_3}$} & \textbf{$\boldsymbol{P_4}$} &
            \textbf{$\boldsymbol{U_2}$} \\
        
            \midrule
            
            \textbf{\#1} & 0.36 & 0.72 & 0.08 & 0.60 & 0.12 & 0.16 & 0.16 
                         & 0.40 & 0.68 & 0.16 & 0.28 & 0.00 & 0.08 & 0.12 \\
            
            \textbf{\#2} & 0.56 & 0.76 & 0.28 & 0.76 & 0.32 & 0.20 & 0.32 
                         & 0.48 & 0.68 & 0.12 & 0.16 & 0.08 & 0.24 & 0.16 \\

            \cmidrule(lr){1-15}
            
            \textbf{Avg.} & 0.46 & 0.74 & \textbf{0.18} & 0.68 & 0.22 & \textbf{0.18} & 0.24 
                          & 0.44 & 0.68 & 0.14 & 0.22 & \textbf{0.04} & 0.16 & 0.14 \\
            
            \bottomrule
        \end{tabular}
    \end{threeparttable}
    \vspace{-1.5ex}
\end{table}

We conduct a user study to evaluate the human imperceptibility of \ours{}. Following prior works~\cite{2023-BADCODE, 2023-CodeMark, 2024-Stealthy-Backdoor-Attack-for-Code-Models}, we construct a human review dataset by mixing watermarked and bare samples at a 1:3 ratio. For each watermark, we randomly select 20 code snippets, five of which contain the watermark. We recruit 4 undergraduate and 6 graduate students in computer science, all of whom have 1-5 years of development experience and are familiar with both Python and Java. We randomly divide them into two groups of five (2 undergraduates and 3 graduates) to separately review the Python and Java datasets. Each participant completes two rounds of annotation. In the first round, only the existence of potential watermarks is disclosed, while in the second round, more watermark implementation details, such as the presence of watermarks involving variable name concatenation, are provided. We quantify the human imperceptibility using suspicious rates~\cite{2023-CodeMark}, defined as the proportion of watermarked samples marked as suspicious by human participants. All the materials for this human study are available in our repository~\cite{PuzzleMark}.

The results of the human study are presented in Table~\ref{tab:rq2_human_study}. CoProtector~\cite{2022-CoProtector} exhibits the weakest imperceptibility, with its sentence-level watermarks $S_1$ and $S_2$ reaching average suspicious rates (ASRs) of 0.74 and 0.68, respectively. Even the more subtle word-level watermarks $W_1$ and $W_2$ result in ASRs exceeding 0.44. In contrast, both \ours{} and CodeMark~\cite{2023-CodeMark} demonstrate substantially stronger imperceptibility. For instance, $B_1$ and $P_2$ achieve the lowest ASR of 0.18 on the Python dataset, while $P_3$ achieves an ASR of 0.04 on the Java dataset. Notably, the $B_2$ of CodeMark results in a substantially higher ASR than the other three implementations, highlighting that the imperceptibility of CodeMark is highly dependent on the specific design of its SPT rules. By comparison, benefiting from its simple design, \ours{} demonstrates more consistent and reliable performance. Further analysis reveals that providing more watermark implementation details leads to an average increase in suspicious rate of 0.14 for Python and 0.03 for Java in Round~\#2 compared to Round~\#1, indicating that more information facilitates watermark identification. Nevertheless, \ours{} maintains strong human imperceptibility even when implementation details are disclosed, with all suspicious rates remaining below 0.32.

\begin{table}[!t]
    \centering
    \scriptsize
    \vspace{-1ex}
    \caption{The FPR and Recall of four automated watermark detection methods.}
    \vspace{-0.8ex}
    \label{tab:rq2_watermark_detection}
    
    \begin{threeparttable}
        \begin{tabular}{ccccccccccc}
            \toprule
        
            \multirow{2}{*}{\textbf{Language}} & 
            \multirow{2}{*}{\textbf{WaterMark}} & 
            \multirow{2}{*}{\textbf{WID.}} & 
            \multicolumn{2}{c}{\textbf{SS}} & 
            \multicolumn{2}{c}{\textbf{AC}} & 
            \multicolumn{2}{c}{\textbf{KillBadCode}} & 
            \multicolumn{2}{c}{\textbf{DeCoMa}} \\
        
            \cmidrule(lr){4-5} \cmidrule(lr){6-7}
            \cmidrule(lr){8-9} \cmidrule(lr){10-11} 
            
             & & &
            \textbf{FPR} & \textbf{Recall} & \textbf{FPR} & \textbf{Recall} &
            \textbf{FPR} & \textbf{Recall} & \textbf{FPR} & \textbf{Recall} \\
        
            \midrule
            
            \multirow{15}{*}{\textbf{Python}} & \textbf{Bare} & - & 
            7.14 & - & 49.75 & - & 40.96 & - & 39.05 & - \\

            \cmidrule(lr){2-11}
            
             & \multirow{3}{*}{\textbf{CoProtector}} & $\boldsymbol{W_1}$ &
            6.92 & 11.29 & 21.03 & 54.52 & 28.89 & 100.00 & 38.01 & 100.00 \\
            
             & & $\boldsymbol{S_1}$ &
            7.11 & 7.77 & 19.32 & 17.39 & 25.06 & 100.00 & 39.09 & 100.00 \\

            \cmidrule(lr){3-11}

             & & \textbf{Average} &
            7.02 & 9.53 & 20.18 & 35.96 & 26.98 & 100.00 & 38.55 & 100.00 \\
        
            \cmidrule(lr){2-11}

             & \multirow{4}{*}{\textbf{CodeMark}} & $\boldsymbol{B_1}$ &
            7.01 & 20.52 & 19.23 & 42.88 & 40.72 & 64.67 & 38.65 & 97.99 \\
            
             & & $\boldsymbol{B_2}$ &
            7.02 & 11.66 & 47.90 & 64.23 & 40.23 & 66.81 & 37.79 & 99.21 \\

             & & $\boldsymbol{M_1}$ &
            6.90 & 13.61 & 48.08 & 57.71 & 40.02 & 65.72 & 37.77 & 98.82 \\

            \cmidrule(lr){3-11}
            
             & & \textbf{Average} &
            6.98 & 15.26 & \textbf{38.40} & 54.94 & 40.32 & 65.73 & 38.07 & 98.67 \\

            \cmidrule(lr){2-11}

             & \multirow{4}{*}{\textbf{\ours{}}} & $\boldsymbol{P_1}$ &
            7.19 & 3.01 & 19.95 & 9.95 & 42.06 & 29.29 & 39.11 & 33.24 \\
            
             & &$\boldsymbol{P_2}$ &
            7.18 & 3.96 & 19.47 & 13.37 & 41.08 & 29.87 & 39.15 & 27.11 \\

             & & $\boldsymbol{U_1}$ &
            7.23 & 5.46 & 40.65 & 40.44 & 41.45 & 32.06 & 39.41 & 28.41 \\

            \cmidrule(lr){3-11}

             & & \textbf{Average} &
            \textbf{7.20} & \textbf{4.14} & 26.69 & \textbf{21.25} & \textbf{41.53} & \textbf{30.41} & \textbf{39.22} & \textbf{29.59} \\

            \cmidrule(lr){1-11}
            
            \multirow{15}{*}{\textbf{Java}} & \textbf{Bare} & - &
            7.14 & - & 17.47 & - & 27.71 & - & 33.82 & - \\

            \cmidrule(lr){2-11}
            
             & \multirow{3}{*}{\textbf{CoProtector}} & $\boldsymbol{W_2}$ &
            6.69 & 15.81 & 29.24 & 34.20 & 25.86 & 100.00 & 32.10 & 100.00 \\
            
             & & $\boldsymbol{S_2}$ &
            7.15 & 6.92 & 26.29 & 20.46 & 17.90 & 100.00 & 32.71 & 100.00 \\

            \cmidrule(lr){3-11}

             & & \textbf{Average} &
            6.92 & 11.37 & \textbf{27.77} & 27.33 & 21.88 & 100.00 & 32.91 & 100.00 \\
        
            \cmidrule(lr){2-11}

             & \multirow{4}{*}{\textbf{CodeMark}} & $\boldsymbol{B_3}$ &
            7.08 & 12.97 & 22.31 & 13.52 & 27.37 & 60.49 & 33.57 & 99.77 \\
            
             & & $\boldsymbol{B_4}$ &
            7.10 & 16.70 & 22.21 & 15.94 & 27.59 & 56.23 & 33.59 & 99.90 \\

             & & $\boldsymbol{M_2}$ &
            7.04 & 14.08 & 22.33 & 14.34 & 27.25 & 59.11 & 33.38 & 99.80 \\

            \cmidrule(lr){3-11}

             & & \textbf{Average} &
            7.07 & 14.58 & 22.28 & \textbf{14.60} & \textbf{27.40} & 58.61 & 33.51 & 99.82 \\

            \cmidrule(lr){2-11}

             & \multirow{4}{*}{\textbf{\ours{}}} & $\boldsymbol{P_3}$ &
            7.16 & 5.08 & 28.61 & 18.38 & 27.71 & 35.04 & 33.88 & 31.21 \\
            
             & & $\boldsymbol{P_4}$ &
            7.15 & 6.07 & 28.19 & 18.84 & 27.73 & 23.90 & 33.82 & 24.23 \\

             & & $\boldsymbol{U_2}$ &
            7.23 & 5.49 & 20.61 & 22.76 & 26.34 & 21.84 & 33.85 & 27.52 \\

            \cmidrule(lr){3-11}

             & & \textbf{Average} &
            \textbf{7.18} & \textbf{5.55} & 25.80 & 19.99 & 27.26 & \textbf{26.93} & \textbf{33.85} & \textbf{27.65} \\
            
            \bottomrule
        \end{tabular}
    \end{threeparttable}
    \vspace{-1ex}
\end{table}

To evaluate the machine imperceptibility of \ours{}, we compare its watermark evasion performance with CoProtector~\cite{2022-CoProtector} and CodeMark~\cite{2023-CodeMark} against four automated detection methods: SS~\cite{2018-spectral-signatures}, AC~\cite{2019-activation-clustering}, KillBadCode~\cite{2025-KillBadCode} and DeCoMa~\cite{2025-DeCoMa}. Since SS and AC require access to latent representations, we fine-tune a DeepSeek-Coder model for each watermark dataset. As shown in Table~\ref{tab:rq2_watermark_detection}, the experimental results demonstrate that both SS and AC perform poorly in watermark detection, achieving either extremely low recall or only moderate recall at the cost of a high FPR. KillBadCode, as a state-of-the-art backdoor defense, achieves strong performance on CoProtector and CodeMark, with average recall of 100\% and around 60\%, respectively. However, its effectiveness is significantly limited on \ours{}, where it only detects 30.41\% of Python and 26.93\% of Java watermarked samples on average. DeCoMa exhibits the strongest watermark detection capability, achieving nearly 100\% recall on both CoProtector and CodeMark. Nevertheless, \ours{}, owing to its unique concatenation pattern, enables both fixed-trigger and universal watermarks to effectively evade DeCoMa, with average recall dropping to just 29.59\% for Python and 27.65\% for Java. These findings indicate that existing watermark detection methods cannot effectively identify the watermarks of \ours{}, demonstrating its strong machine imperceptibility.

\begin{tcolorbox}[size=title]
{\textbf{Answer to RQ2:}}
\ours{} achieves imperceptibility for manual inspection comparable to CodeMark, and exhibits significantly higher machine imperceptibility against various watermark detection methods compared to all baselines.
\end{tcolorbox}

\subsubsection{RQ3: How robust is \ours{} against removal and dilution attacks?}
\label{subsubsec:rq3}
\

\begin{figure}[t]
    \centering
    
    \begin{minipage}[c]{\textwidth}
        \begin{table}[H]
    \centering
    \scriptsize
    \caption{The $p$-values of \ours{} under four removal attacks on DeepSeek-Coder.}
    \vspace{-1ex}
    \label{tab:rq3_removal_attack}
    
    \begin{threeparttable}
        \begin{tabular}{ccccccc}
            \toprule
        
            \multirow{2}{*}{\textbf{Attack}} & 
            \multicolumn{3}{c}{\textbf{Python}} & 
            \multicolumn{3}{c}{\textbf{Java}} \\
        
            \cmidrule(lr){2-4} \cmidrule(lr){5-7} & 
            \textbf{$\boldsymbol{P_1}$} & \textbf{$\boldsymbol{P_2}$} & \textbf{$\boldsymbol{U_1}$} &
            \textbf{$\boldsymbol{P_3}$} & \textbf{$\boldsymbol{P_4}$} & \textbf{$\boldsymbol{U_2}$} \\
        
            \midrule
            
            \textbf{SS} & 
            \graycell{}7.32E-145 & \graycell{}2.27E-185 & \graycell{}4.88E-170 & 
            \graycell{}2.42E-235 & \graycell{}9.52E-234 & \graycell{}1.44E-237 \\

            \textbf{AC} & 
            \graycell{}6.43E-186 & \graycell{}2.21E-170 & \graycell{}3.42E-122 & 
            \graycell{}2.59E-210 & \graycell{}1.71E-225 & \graycell{}1.06E-243 \\

            \textbf{KillBadCode} & 
            \graycell{}1.13E-141 & \graycell{}5.66E-180 & \graycell{}1.38E-156 & 
            \graycell{}2.65E-226 & \graycell{}1.44E-221 & \graycell{}1.81E-204 \\

            \textbf{DeCoMa} & 
            \graycell{}3.54E-167 & \graycell{}2.62E-169 & \graycell{}2.99E-157 & 
            \graycell{}1.04E-224 & \graycell{}1.10E-211 & \graycell{}2.97E-247 \\
            
            \bottomrule
        \end{tabular}
    \end{threeparttable}
\end{table}
    \end{minipage}

    \begin{minipage}[c]{\textwidth}
        \begin{table}[H]
    \centering
    \scriptsize
    \caption{The $p$-values of \ours{} under various watermarking rates on DeepSeek-Coder.}
    \vspace{-1ex}
    \label{tab:rq3_dilution_attack}
    
    \begin{threeparttable}
        \begin{tabular}{ccccccc}
            \toprule
        
            \multirow{2}{*}{\textbf{Rate}} & 
            \multicolumn{3}{c}{\textbf{Python}} & 
            \multicolumn{3}{c}{\textbf{Java}} \\
        
            \cmidrule(lr){2-4} \cmidrule(lr){5-7} & 
            \textbf{$\boldsymbol{P_1}$} & \textbf{$\boldsymbol{P_2}$} & \textbf{$\boldsymbol{U_1}$} &
            \textbf{$\boldsymbol{P_3}$} & \textbf{$\boldsymbol{P_4}$} & \textbf{$\boldsymbol{U_2}$} \\
        
            \midrule
            
            \textbf{1\%} & 
            \graycell{}1.81E-204 & \graycell{}2.38E-163 & \graycell{}2.01E-72 & 
            \graycell{}6.15E-230 & \graycell{}1.26E-219 & \graycell{}5.77E-133 \\

            \textbf{0.5\%} & 
            \graycell{}3.71E-152 & \graycell{}1.20E-152 & \graycell{}5.54E-25 & 
            \graycell{}8.67E-201 & \graycell{}8.40E-175 & \graycell{}1.79E-82 \\

            \textbf{0.2\%} & 
            \graycell{}4.76E-74 & \graycell{}3.74E-86 & \graycell{}1.57E-06 & 
            \graycell{}3.79E-168 & \graycell{}5.35E-150 & \graycell{}1.76E-87 \\

            \textbf{0.1\%} & 
            \graycell{}5.37E-50 & \graycell{}5.73E-58 & 1.00E+00 & 
            \graycell{}2.88E-99 & \graycell{}9.63E-175 & \graycell{}1.60E-37 \\
            
            \bottomrule
        \end{tabular}
    \end{threeparttable}
\end{table}
    \end{minipage}

    \vspace{-1ex}

    \Description{}
\end{figure}

Since DeCoMa~\cite{2025-DeCoMa} achieves nearly 100\% recall on both CoProtector~\cite{2022-CoProtector} and CodeMark~\cite{2023-CodeMark}, almost all watermarked samples can be simply removed from the dataset to undermine watermark verifiability. We systematically evaluate the resilience of \ours{} against such removal attacks based on four existing automatic detection methods. Specifically, for each watermarked dataset, we construct four purified subsets by separately removing all suspicious samples identified by each detection method. These subsets are then used to fine-tune DeepSeek-Coder models, followed by Fisher's exact test to verify the presence of watermarks. As shown in Table~\ref{tab:rq3_removal_attack}, the experimental results demonstrate the strong robustness of \ours{}, which consistently preserves watermark verifiability (i.e., $p < 0.05$) against all four removal attacks.

In practice, data collectors often merge datasets from diverse sources, most of which are unwatermarked. Introducing substantial amounts of clean data reduces the proportion of watermarked samples, which increases the risk of dilution attacks and potentially undermines watermark effectiveness. We simulate such attacks by setting maximum watermark embedding rates to 1\%, 0.5\%, 0.2\%, and 0.1\%, fine-tuning DeepSeek-Coder models, and applying Fisher's exact test to their outputs for watermark detection. As shown in Table~\ref{tab:rq3_dilution_attack}, the watermark validation becomes statistically less significant as the embedding rate decreases. Notably, the $U_1$ watermark fails when the embedding rate drops to 0.1\%, as universal watermarks generally require higher rates due to their subtle patterns. All other watermarks remain effective even at 0.1\%, highlighting the strong robustness of \ours{} against dilution attacks. It is worth noting that diluting a 1\% watermark rate to 0.1\% in CSN-Python, a dataset with 0.41 million samples, requires collecting 3.69 million additional clean samples, which further underscores the inherent challenges of dilution attacks in practice.

\begin{tcolorbox}[size=title]
{\textbf{Answer to RQ3:}}
PuzzleMark consistently maintains watermark verifiability under both removal and dilution attacks, demonstrating strong robustness against adverse attacks.
\end{tcolorbox}

\subsubsection{RQ4: How does \ours{} perform under different settings, including the suitability score threshold, feature projection methods and training hyperparameters?}
\label{subsubsec:rq4}
\

\begin{figure}[t]
    \centering

    \begin{subfigure}
        \centering
        \begin{minipage}[t]{0.64\linewidth}
            \centering
            \includegraphics[width=\linewidth, trim=10 10 10 10, clip, keepaspectratio]{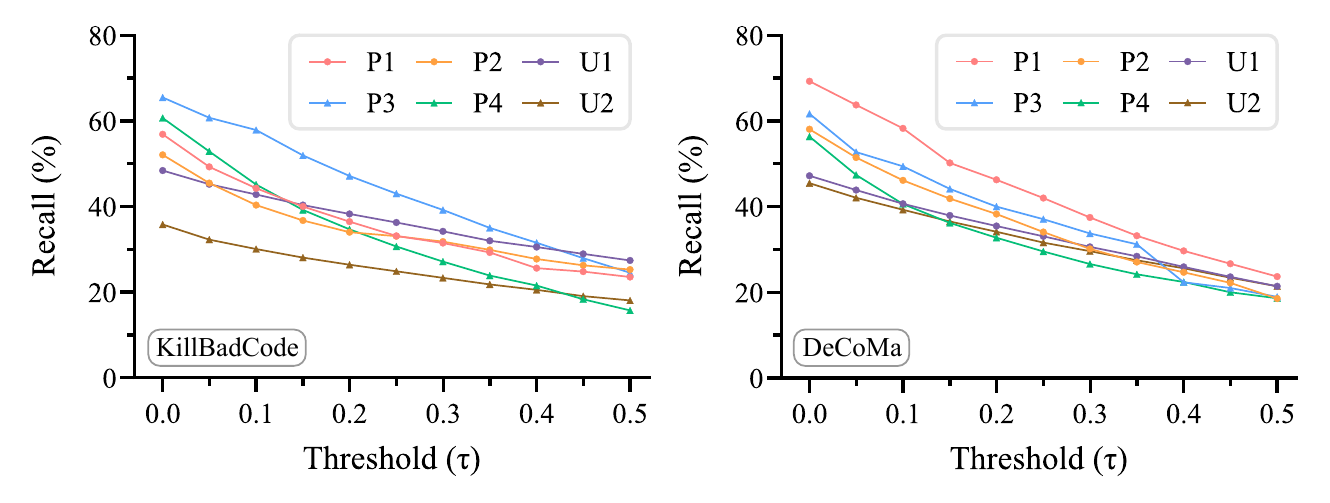}
            \vspace{-3.5ex}
            \caption{The Recall of DeCoMa~\cite{2025-DeCoMa} and KillBadCode~\cite{2025-KillBadCode} against \ours{} under different suitability score threshold $\tau$.}
            \label{fig:threshold-recall}
        \end{minipage}
        \hfill
        \begin{minipage}[t]{0.335\linewidth}
            \centering
            \includegraphics[width=\linewidth, trim=10 8 10 15, clip,  keepaspectratio]{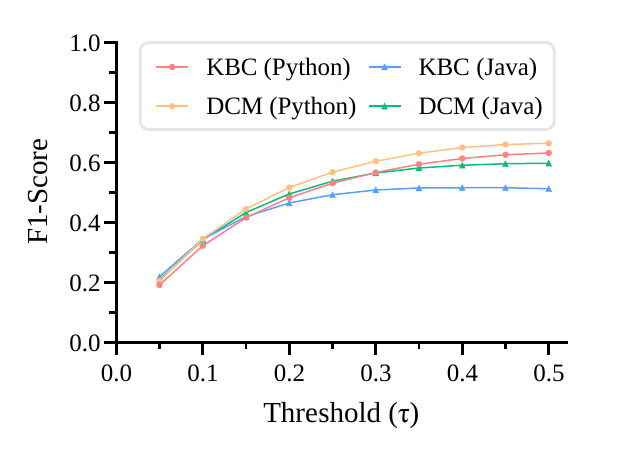}
            \vspace{-3.5ex}
            \caption{The F1-scores for identifying detectable carriers under different $\tau$.}
            \label{fig:threshold-f1}
        \end{minipage}
    \end{subfigure}
    
    \begin{subfigure}
        \centering
        \includegraphics[width=\linewidth, trim=15 15 15 0, clip, keepaspectratio]{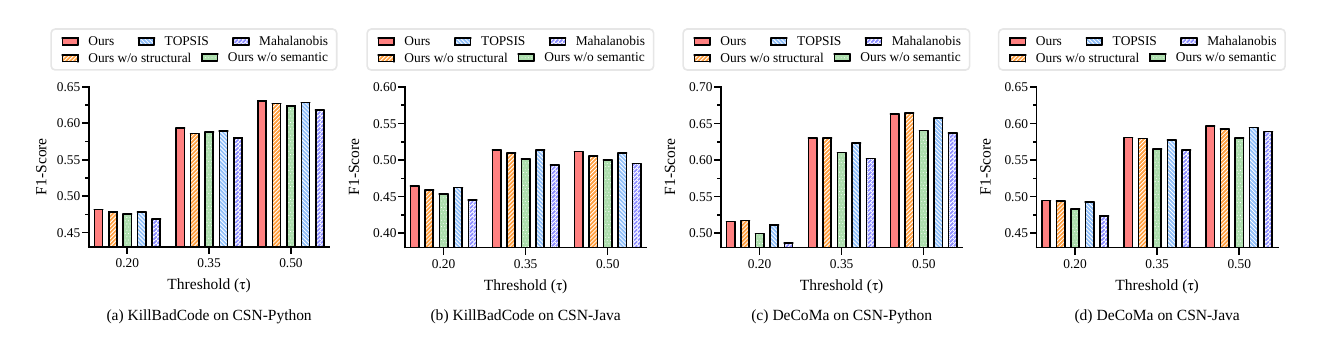}
        \vspace{-3.5ex}
        \caption{The F1-scores of \ours{} for identifying detectable carriers under different feature projection methods and feature ablation settings.}
        \label{fig:projection-ablation}
    \end{subfigure}

    \vspace{-1ex}

    \begin{subfigure}
        \centering
        \begin{minipage}[c]{0.60\textwidth}
            \begin{table}[H]
    \centering
    \scriptsize
    \caption{The $p$-values of \ours{} under different training hyperparameters on DeepSeek-Coder (DSC).}
    \vspace{-1.5ex}
    \label{tab:rq4_hyperparam}
    
    \begin{threeparttable}
        \setlength{\tabcolsep}{4pt}
        \begin{tabular}{ccccccc}
            \toprule
        
            \multirow{2}{*}{\textbf{WID.}} &
            \multicolumn{3}{c}{\textbf{Learning Rate}} & 
            \multicolumn{3}{c}{\textbf{Epoch}} \\
        
            \cmidrule(lr){2-4} \cmidrule(lr){5-7} &
            \textbf{1e-3} & \textbf{1e-4} & \textbf{1e-5} &
            \textbf{1} & \textbf{5} & \textbf{10} \\
        
            \midrule
            
            \textbf{$\boldsymbol{P_1}$} & 
            \graycell{}3.94E-140 & \graycell{}1.81E-204 & \graycell{}6.83E-145 & 
            \graycell{}4.89E-183 & \graycell{}3.77E-200 & \graycell{}1.81E-204 \\
            
            \textbf{$\boldsymbol{P_2}$} & 
            \graycell{}2.39E-164 & \graycell{}2.38E-163 & \graycell{}6.65E-143 & 
            \graycell{}7.86E-167 & \graycell{}1.08E-211 & \graycell{}2.38E-163 \\

            \textbf{$\boldsymbol{U_1}$} & 
            \graycell{}4.30E-121 & \graycell{}6.35E-156 & \graycell{}6.62E-136 & 
            \graycell{}1.15E-169 & \graycell{}1.69E-181 & \graycell{}6.35E-156 \\
            
            \bottomrule
        \end{tabular}
    \end{threeparttable}
    \vspace{-1ex}
\end{table}
        \end{minipage}
        \hfill
        \begin{minipage}[c]{0.37\textwidth}
            \begin{table}[H]
    \centering
    \scriptsize
    \vspace{-0.5ex}
    \caption{The $p$-values of \ours{} across different-scaled projects on DSC.}
    \vspace{-1.5ex}
    \label{tab:rq5_project_protection}
    
    \begin{threeparttable}
        \setlength{\tabcolsep}{4pt}
        \begin{tabular}{cccc}
            \toprule
        
            \multirow{2}{*}{\textbf{WID.}} &
            \multicolumn{3}{c}{\textbf{Project}} \\
        
            \cmidrule(lr){2-4} &
            \textbf{Osmnx} & \textbf{Spark} & \textbf{Pandas} \\
        
            \midrule
            
            \textbf{$\boldsymbol{P_1}$} & 
            1.00E+00 & \graycell{}8.73E-46 & \graycell{}4.93E-93 \\
            
            \textbf{$\boldsymbol{P_2}$} & 
            \graycell{}2.91E-02 & \graycell{}9.88E-60 & \graycell{}3.04E-77 \\

            \textbf{$\boldsymbol{U_1}$} & 
            1.00E+00 & \graycell{}1.30E-23 & \graycell{}4.57E-15 \\
            
            \bottomrule
        \end{tabular}
    \end{threeparttable}
    \vspace{-1ex}
\end{table}
        \end{minipage}
    \end{subfigure}

    \vspace{-1ex}

    \Description{}
\end{figure}

To evaluate the impact of the suitability score threshold $\tau$ in \ours{}, we conduct experiments varying $\tau$ from 0 to 0.5 in steps of 0.05, using the two most effective watermark detection methods, KillBadCode (KBC)~\cite{2025-KillBadCode} and DeCoMa (DCM)~\cite{2025-DeCoMa}. We observe from Figure~\ref{fig:threshold-recall} that increasing $\tau$ results in a significant drop in watermark recall, as higher thresholds filter out more unnatural carriers and thus enhance the imperceptibility of \ours{}. However, an excessively high $\tau$ reduces the number of samples with naturally occurring trigger prefixes, forcing \ours{} to introduce more code perturbations and increasing the risk of exposing the watermark patterns. To address this trade-off, we further evaluate the F1-scores for identifying detectable carriers under different $\tau$ values, where code snippets marked as suspicious by KillBadCode or DeCoMa on the bare dataset are treated as positive samples. This process produced four datasets, derived from two detection methods and two programming languages. As illustrated in Figure~\ref{fig:threshold-f1}, the F1-score plateaus at around 0.35, leading us to select $\tau = 0.35$ as the optimal trade-off in our experiments.

Beyond the suitability score threshold, the choice of feature projection method can also impact the performance of \ours{}. To investigate this, we compare our approach, which projects features onto principal component directions, with two baseline methods: TOPSIS, which is widely used in multiple attribute decision making~\cite{1995-MADM}, and Mahalanobis distance, which measures how far a sample deviates from the mean of a distribution. Additionally, we perform feature ablation by separately removing the three structural features and four semantic features, resulting in two further baselines. We conduct experiments on the same four datasets as mentioned above. The results, as shown in Figure~\ref{fig:projection-ablation}, indicate that \ours{} is generally insensitive to the choice of feature projection method. This can be attributed to our carefully selected features in Section~\ref{subsubsec:feature-selection}. Overall, our method achieves the highest F1-score in all 12 cases, outperforming both TOPSIS and Mahalanobis distance, demonstrating its effectiveness. The ablation results further show that removing structural features causes a slight drop, while removing semantic features leads to a more significant decline, highlighting the greater importance of semantic features, yet the combination of both yields the best performance.

Furthermore, attackers may attempt to manipulate training hyperparameters in an effort to compromise watermark verifiability. To assess whether \ours{} remains robust under varying training conditions, we conduct experiments on DeepSeek-Coder using the CSN-Python, systematically altering two critical hyperparameters: learning rate and training epochs. As shown in Table~\ref{tab:rq4_hyperparam}, \ours{} consistently maintains strong verification performance across all training configurations, highlighting its strong robustness against variations in training hyperparameters.

\begin{tcolorbox}[size=title]
{\textbf{Answer to RQ4:}}
Experimental results show that increasing the suitability score threshold $\tau$ effectively reduces watermark detection rates, with the best trade-off achieved at $\tau = 0.35$. Our feature projection method attains optimal performance in carrier selection, and ablation studies further validate the effectiveness of the selected complexity features. Additionally, \ours{} exhibits strong robustness against hyperparameter tuning.
\end{tcolorbox}

\subsubsection{RQ5: How scalable is \ours{} to project-level copyright protection?}
\label{subsubsec:rq5}
\

Although \ours{} is a watermarking technique designed for large-scale code datasets, practical needs for ownership verification also extend to individual developers and organizational projects~\cite{2022-CoProtector}. We systematically evaluate the extensibility of \ours{} for project-level copyright protection. Specifically, we select three Python projects of different scales from CSN-Python: gboeing/osmnx, apache/spark, and pandas-dev/pandas, with approximately 100, 400, and 1000 watermarkable functions respectively. \ours{} restricts watermark embedding exclusively to the target project's codebase, leaving the remainder of the dataset intact. We fine-tune DeepSeek-Coder and employ Fisher's exact test for watermark verification. Experimental results are shown in Table~\ref{tab:rq5_project_protection}. Even when only a single project within a large corpus is watermarked, \ours{} remains highly effective, demonstrating its potential for safeguarding project-level intellectual property. However, its performance on smaller projects (e.g., Osmnx) is less ideal. This is expected, as both backdoor poisoning~\cite{2022-you-see-what-I-want-you-to-see, 2023-BADCODE, 2024-Stealthy-Backdoor-Attack-for-Code-Models, 2024-TrojanPuzzle} and watermarking~\cite{2022-CoProtector, 2023-CodeMark} rely on a sufficient poisoning rate. An extremely low rate provides insufficient signal for the model to discern and learn the backdoor patterns, leading to implantation failure. This highlights the inherent challenges in project-level copyright protection. Nevertheless, \ours{} remains capable of effectively protecting the majority of projects with moderate scale, underscoring its practical utility.

\begin{tcolorbox}[size=title]
{\textbf{Answer to RQ5:}}
\ours{} is also capable of protecting project-level intellectual property, albeit with some limitations in smaller projects.
\end{tcolorbox}
\section{Discussion}
\label{sec:discussion}

\noindent
\textbf{Generalizability of \ours{}.}
Following CodeMark~\cite{2023-CodeMark}, this paper primarily focuses on code completion, a representative code-code task. However, given that the threat model assumes we do not know what models attackers may train on the dataset, evaluating only this task is insufficient. Although \ours{} preserves textual information such as comments during watermarking, it does not fully exploit these elements, making it currently unsuitable for text-code tasks (e.g., code generation) and code-text tasks (e.g., code summarization). Nevertheless, this does not imply that \ours{} lacks generalizability. In fact, the concatenation pattern proposed by \ours{} are not limited to variable names and have potential for application to token concatenation in natural language text. In this work, we have conducted preliminary exploration of the feasibility of this pattern within code context and demonstrated its advantages over the traditional co-occurrence pattern. We plan to further investigate the implementation of \ours{} in other tasks in future work. Moreover, although evaluation was conducted on datasets in only two programming languages, the core design of \ours{} relies solely on variable renaming within ASTs, which is fundamentally language-agnostic. As long as the naming conventions and parsers of target languages are provided, \ours{} can be automatically adapted to any programming language.

\noindent
\textbf{Robustness of \ours{}.}
Since the watermark in \ours{} is embedded in the form of variable names, attackers may attempt to eliminate the watermark through code obfuscation. However, the fundamental purpose of code obfuscation is to reduce code readability and analyzability. Existing obfuscation tools, such as Anubis~\cite{2022-Anubis} and Pyarmor~\cite{2025-Pyarmor}, are difficult to precisely control in terms of obfuscation intensity~\cite{2024-CodeBreaker}. Consequently, while these tools may disrupt the watermark, they inevitably cause a significant decline in dataset quality. Even if attackers design targeted obfuscation strategies specifically for variable names, since they cannot know which samples are watermarked, they must obfuscate the entire dataset, globally erasing naming semantics. This is highly detrimental, as variable name semantics are known to be crucial for both code understanding~\cite{2024-how-effectively-do-clms-understand-poor-readability-code, 2023-two-sides-of-the-same-coin} and generation~\cite{2023-recode}. Thus, attackers must carefully balance data quality and obfuscation strength. Furthermore, although leveraging LLMs for code rewriting appears to be a feasible attack strategy, applying such rewriting across an entire dataset would require extensive time and financial resources, making it impractical~\cite{2025-DeCoMa}. A potential approach is to first use automated tools, such as DeCoMa~\cite{2025-DeCoMa}, to detect candidate watermarked samples, and then perform rewriting attacks on them. However, existing watermark detection methods have limited effectiveness against \ours{}. These limitations mean that attackers must design more comprehensive or costly attack strategies, making it significantly more difficult and less economical to steal a dataset.

\noindent
\textbf{Synergy between Components.}
The plug-and-play nature of our carrier selection allows it to extend to watermarks like CoProtector~\cite{2022-CoProtector}, whereas CodeMark~\cite{2023-CodeMark} remains incompatible due to its rigid carrier constraints. However, experiments show that even when combined with carrier selection, CoProtector fails to evade DeCoMa~\cite{2025-DeCoMa}, yielding a detection rate of 100\%. This stark contrast confirms that the code naturalness ensured by carrier selection and the stealth inherent in adaptive watermarking are mutually reinforcing in \ours{}.

\noindent
\textbf{Additional Time Cost.}
The computational overhead of \ours{} primarily lies in carrier selection, which takes about 30 minutes for 400K samples. The subsequent watermarking process requires only 5 minutes, comparable to CodeMark\cite{2023-CodeMark} and slightly longer than the 2 minutes of CoProtector\cite{2022-CoProtector}. Although carrier selection introduces additional overhead, it is an offline, one-time procedure, and our experiments consistently demonstrate its importance for watermark robustness and reliability. Therefore, the extra time spent on carrier selection is acceptable in practice.

\noindent
\textbf{Unintended Activation of Universal Watermark.}
The dynamic variability of the universal watermark may increase the risk of unintended activation. To quantify this risk, we simulate 1,000 code-completion requests on the DeepSeek-Coder models fine-tuned with the $U_1$ and $U_2$ watermarks. The observed unintended activation rates are only 0.8\% and 0.1\%, respectively. While the universal watermark theoretically broadens the trigger space and raises the likelihood of unintended activation, its actual impact on user experience remains limited in practice.
\section{Related Work}
\label{sec:related_work}

\noindent
\textbf{Source Code Watermarking.}
The open-source paradigm has made plagiarism and unauthorized copying of code increasingly prevalent. Existing techniques such as clone detection~\cite{2019-Learning-Based-Recursive-Aggregation-of-AST-for-Code-Clone-Detection} and authorship attribution~\cite{2018-DL-CAIS, 2019-Code-Authorship-Attribution-Methods-and-Challenges} fall short in establishing clear ownership and thus lack the evidentiary strength for legal protection. Source code watermarking aims to trace code provenance, enabling original authors to claim ownership in cases of dispute. SrcMarker~\cite{2024-SrcMarker} is currently the only deep-learning source code watermarking system. It utilizes a learnable encoder-decoder architecture to embed unobtrusive ID bit strings into source code through semantics-preserving transformations. Unlike dataset watermarking, which embeds verifiable signals in models to trace unauthorized dataset usage, source code watermarking focuses on code provenance for ownership verification.

\noindent
\textbf{Code Model Watermarking.}
LLM-based commercial tools like GitHub Copilot~\cite{2022-GitHub-Copilot} have become essential in modern software development, but their training is highly resource-intensive. To prevent unauthorized copying or theft, model watermarking techniques have emerged. SWEET~\cite{2024-Who-Wrote-this-Code} is the first watermarking method for code generation LLMs, based on WLLM~\cite{2023-Watermark-for-LLMs}. It selectively applying red-green rules only to high-entropy tokens, enabling reliable detection of machine-generated code with minimal impact on code quality. ModMark~\cite{2025-ModMark}, the first watermarking method for code summarization models, embeds verifiable and harmless watermarks by fine-tuning the tokenizer. While model watermarking and dataset watermarking safeguard different assets, both are essential for deep learning intellectual property protection.

\noindent
\textbf{Code Dataset Poisoning.}
There is a close relationship between poisoning and watermarking in code datasets, as both involve embedding specific signals into the data to influence model behavior. Data poisoning aims to degrade model performance (untargeted poisoning) or to induce models to produce erroneous or even malicious outputs in specific contexts (targeted poisoning). Poisoning techniques may also serve dataset protection purposes, as exemplified by CoProtector~\cite{2022-CoProtector}, which introduces four untargeted poisoning strategies that can significantly impair model performance and thus intimidate potential dataset thieves. However, such approaches violate the harmlessness requirement of code dataset protection~\cite{2023-CodeMark}, limiting their practical applicability. In contrast, dataset watermarking embeds benign and verifiable signals into datasets, offering a more practical and effective means of intellectual property protection.

\noindent
\textbf{Code Dataset Watermarking.}
Dataset watermarking techniques are designed to prevent unauthorized use of datasets for model training. CoProtector~\cite{2022-CoProtector} is the first watermarking method specifically designed for code datasets, embedding watermarks through fixed variable renaming and dead code insertion. CodeMark~\cite{2023-CodeMark} further enhances the imperceptibility of watermarks by introducing semantics-preserving transformations. However, recent work, DeCoMa~\cite{2025-DeCoMa}, has revealed the limitations of both methods. To address these challenges, we propose \ours{} as a novel solution for code dataset watermarking.
\section{Conclusion}
\label{sec:conclusion}

In this paper, we propose \ours{}, a simple yet highly robust code dataset watermarking technique designed to effectively prevent unauthorized training of NCCMs. \ours{} introduces a novel concatenation pattern to replace the traditional co-occurrence pattern for watermarking, and implements two watermarking strategies via variable name concatenation. Due to the critical role of code carriers in watermark robustness, \ours{} further optimizes carrier selection to reduce the risk of watermark exposure. Experimental results demonstrate that \ours{} fulfills all practical requirements for watermarking, including effectiveness, harmlessness, imperceptibility, and robustness, making it a promising tool for protecting code datasets.

\section{Data Availability}
\textit{Our source code and experimental data are available at~\cite{PuzzleMark}.}

\section*{Acknowledgement}
This work is partially supported by the National Natural Science Foundation of China (62172202), the Major Program of the Natural Science Foundation of Jiangsu Higher Education Institutions of China under Grant No.22KJA520008, the Collaborative Innovation Center of Novel Software Technology and Industrialization, the Priority Academic Program Development of Jiangsu Higher Education Institutions, as well as the National Research Foundation, Singapore, and DSO National Laboratories under the AI Singapore Programme (AISG Award No: AISG4-GC-2023-008-1B).

\bibliographystyle{ACM-Reference-Format}
\bibliography{reference}

\end{document}